\documentclass[10pt,letterpaper,journal]{IEEEtran}
\usepackage{lineno,hyperref}
\usepackage{bm}
\usepackage{mathrsfs}
\usepackage{amsmath,cases}
\usepackage{amssymb}
\usepackage{textcomp}
\usepackage{graphicx}
\usepackage{subfigure}
\usepackage{epstopdf}
\usepackage{xcolor}
\usepackage{overpic}
\modulolinenumbers[5]

\ifCLASSINFOpdf

\else

\fi

\begin{document}

\title{Study of Proximal Normalized Subband Adaptive Algorithm for Acoustic Echo Cancellation}

\author{Gang~Guo,
        ~Yi Yu,~\IEEEmembership{Member,~IEEE},
        ~Rodrigo C. de Lamare,~\IEEEmembership{Senior Member,~IEEE},
        ~Zongsheng~Zheng,~\IEEEmembership{Member,~IEEE},
        ~Lu Lu,~\IEEEmembership{Member,~IEEE},
        ~and Qiangming~Cai
\thanks{This work was partially supported by the National Natural Science Foundation of China (NSFC) (Nos. 61901400, 61901285, and 61801406), and the Sichuan Science and Technology Program (Nos. 20YYJC3709, 2021YFG0253), and the Doctoral Research Fund of Southwest University of Science and Technology in China (No. 19zx7122). Corresponding author: Yi Yu.}

\thanks{Y. Yu and Q. Cai are with School of Information Engineering, Robot Technology Used for Special Environment Key Laboratory of Sichuan Province, Southwest University of Science and Technology, Mianyang, 621010, China (e-mail: yuyi\_xyuan@163.com, qmcai@swust.edu.cn).}

\thanks{G.~Guo is with JingWei HiRain, 5-10F, Block D, Truth Plaza, N0.7 Zhichun Road, Haidian District, Beijing, China. (e-mail:gang.guo@hirain.com).}

\thanks{R. C. de Lamare is with CETUC, PUC-Rio, Rio de Janeiro 22451-900, Brazil, and Department of Electronic Engineering, University of York, York YO10 5DD, U.K. (e-mail: delamare@cetuc.puc-rio.br).}

\thanks{Z. Zheng is with School of Electrical Engineering, Sichuan University, Chengdu, 610065, China. (e-mail: bk20095185@my.swjtu.edu.cn).}

\thanks{L. Lu is with School of Electronics and Information Engineering, Sichuan University, Chengdu, 610065, China (e-mail: lulu19900303@126.com).}
}


\maketitle

\begin{abstract}
In this paper, we propose a novel normalized subband adaptive filter algorithm suited for sparse scenarios, which combines the proportionate and sparsity-aware mechanisms. The proposed algorithm is derived based on the proximal forward-backward splitting and the soft-thresholding methods. We analyze the mean and mean square behaviors of the algorithm, which is supported by simulations. In addition, an adaptive approach for the choice of the thresholding parameter in the proximal step is also proposed based on the minimization of the mean square deviation. Simulations in the contexts of system identification and acoustic echo cancellation verify the superiority of the proposed algorithm over its counterparts.
\end{abstract}


\IEEEpeerreviewmaketitle

\section{Introduction}

\IEEEPARstart{a}{daptive} filtering algorithms have been widely
applied in system identification, echo cancellation (EC), feedback
noise cancellation, and active noise control,
etc~\cite{chen2016generalized,chen2017kernel,lee2009subband,pradhan2017improved,intadap,jio,jidf,sjidf,jiols,jiomimo,wlmwf,wljio,jiodoa,barc,rrbf,rrser,rralr,l1stap,smtvb,smce,dce,damdc,arc,spa,mbdf,bfidd,aaidd,listmtc,msgampmtc,dynovs,locsme,okspme,lrcc,rcoprime}.
In the literature, the least mean square (LMS) algorithm is one of
the widely studied algorithms, owing to its simplicity and
practicality. To overcome the stability of LMS depending on the
maximum eigenvalue of the input correlation matrix, the normalized
LMS (NLMS) algorithm was presented. However, both algorithms will
undergo slow convergence when the input signal is colored (or say,
successive realizations of the input signal are correlated). With
the aim of addressing the problems with such input signals, affine
projection (AP) and recursive least squares (RLS) algorithms that
exhibit fast convergence have been
studied~\cite{sayed2003fundamentals}. Moreover, to obtain low
complexity implementations, several fast AP and RLS versions were
proposed~\cite[Chapter 14]{sayed2003fundamentals},
\cite{yang2018comparative,zakharov2008low}, but most of them are
still prone to numerical instability issues.

Alternatively, subband adaptive filtering (SAF) is an efficient technique to improve the convergence rate in the colored input signal case~\cite{lee2009subband}. In the SAF, the input signal is decomposed into subband signals through the analysis filter bank and then decimated; thus the resulting input signal in each subband is approximately white to update the filter's weights. In~\cite{lee2004improving}, Lee and Gan presented the normalized SAF (NSAF) algorithm from the principle of minimum disturbance, over the multiband structure of SAF. For colored input signals, the NSAF algorithm significantly accelerates the filter weights' convergence in contrast with the NLMS algorithm; also, the former keeps comparable computational complexity with the latter, especially when requiring a long adaptive filter in applications such as EC. In the SAFs, the multiband structure has no aliasing and band edge effects as compared to the conventional structure~\cite{lee2009subband}. Therefore, SAF algorithms founded on the multiband structure have received much attention in the last decade. In~\cite{lee2007delayless}, considering the practical applicability of the NSAF algorithm, the same authors also developed two delayless configurations by computing the estimated output of the system in an auxiliary loop, which overcome the signal delay problem in the original structure caused by the adopted analysis and synthesis filter banks. Since the step-size of the NSAF algorithm determines the tradeoff between convergence and steady-state behaviors, various variable step-size~\cite{ni2009variable,seo2014variable} and combination variants~\cite{ni2010adaptive} were proposed. In~\cite{yang2012improved}, the AP concept was incorporated into the NSAF algorithm to further improve the decorrelation for colored input signals. To reduce the high complexity of the AP type, the authors also provided many implementations with lower complexity in~\cite{yang2015low}.

In adaptive filtering applications, sparse systems are frequently encountered, with the property that the majority of coefficients in the system's impulse response are zero while a few coefficients have values far away from zero. Examples of such systems are network echo channels in the network EC (NEC)~\cite{radecki2002echo}, acoustic echo channels in the acoustic EC (AEC)~\cite{yukawa2008efficient}, the digital transmission channel in high-definition television~\cite{schreiber1995advanced},~and so on. Specifically, the network echo channel has typically a length of 64-128 ms but with an active region in the range of 8-12 ms duration, where this sparsity is due to the presence of bulk delay caused by network propagation, encoding and jitter buffer delays. The acoustic echo channel is the path between microphone and loudspeaker in hands free mobile telephony through which the far-end speaker hears replica of her/his own voice with time lags, where its sparsity determined by many factors such as the loudspeaker-microphone distance~\cite{loganathan2009class,yukawa2008efficient}. As a result, exploiting the sparsity of systems can improve the filter performance. At present, there are two main strategies towards this goal. The first strategy is to introduce the proportionate matrix in the filter's weights update that assigns an individual gain to each filter weight~\cite{benesty2002improved}. It was proposed originally to improve the NLMS performance~\cite{duttweiler2000proportionate}. By combining the merits of both SAF and proportionate idea, a series of proportionate NSAF (PNSAF) algorithms were proposed in~\cite{abadi2009proportionate,abadi2011family}, exhibiting faster convergence than the NSAF algorithm in sparse systems under the same steady-state behavior.
Another sparse adaptive filter is inspired by the compressive sensing framework~\cite{baraniuk2007compressive}. It adds a sparse penalty term based on the $l_p$-norm of the filter weights vector to the original cost function, where $p = 0$, 1, or $0 < p < 1$~\cite{gu2009l,de2014sparsity}. In~\cite{yu2016sparse}, the $l_0$-norm penalty is considered into the NSAF algorithm, thereby obtaining a performance improvement when identifying sparse systems in the colored input scenarios. In~\cite{yu2019sparsity}, based on the $l_1$-norm and reweighted $l_1$-norm penalties, the sparsity-aware NSAF algorithms that outperform the NSAF algorithm were developed and analyzed. It is worth pointing out that the main role of the proportionate scheme is to speed up the convergence, while the sparsity-aware's role is to reduce the steady-state error. In~\cite{pelekanakis2012new,das2016improving,jin2017enhanced,ferreira2016low}, to exploit the sparsity of the underlying system as full as possible, these two strategies were combined in the NLMS algorithm. However, the reasons why the combination of these two strategies provided better results than each strategy individually are not completely clarified, and how to choose the proper sparse penalty parameter is also problem. Furthermore, such sparse technique has been seldom reported in the subband domain. This is also the motivation of this paper, namely, by bringing together the proportionate and sparsity-aware strategies improves the learning performance of the SAF in sparse systems. The main contributions of this paper are as follows:

1) Based on the proximal forward-backward splitting (PFBS) and the soft-thresholding techniques~\cite{yamagishi2011acceleration,jeong2018automatic}, we derive a novel PNSAF algorithm, called the PFBS-PNSAF algorithm.

2) The performance of the PFBS-PNSAF algorithm is analyzed in detail, including the convergence condition, transient state and steady-state behaviors. The analysis results are also supported by simulations.

3) It follows from the performance analysis that, to minimize the mean square deviation (MSD) of the PFBS-PNSAF algorithm, we also derive an adaptive rule to adjust the sparsity-aware thresholding parameter.

4) For the AEC application, the delayless implementation of PFBS-PNSAF is deployed.

At first glance from sparsity-aware strategies, compared with our previous work in~\cite{yu2019sparsity}, this work extends the additional proportionate mechanism for sparse systems. However, unlike~\cite{yu2019sparsity}, the PFBS-PNSAF algorithm is based on the PFBS and the soft-thresholding techniques. Importantly, this work also covers a comprehensive performance analysis for the PNSAF algorithm in terms of convergence condition, transient state and steady-state behaviors (which have not been discussed in detail). Moreover, we consider the delayless SAF in AEC rather than the original SAF structure in~\cite{yu2019sparsity}.

This paper is organized as follows. In Section~II, the multiband structured SAF is described and the NSAF and PNSAF algorithms are revisited. Then, in Section~III, we propose the PFBS-PNSAF algorithm. The mean and mean-square performance of the proposed algorithm are analyzed in Section~IV. The adaptation of $\beta$ is designed in Section~V. In Section~VI, simulation results in both system identification and AEC scenarios are presented. Finally, conclusions are given in Section VII.

\section{Statement of the multiband structured SAF and review of NSAF and PNSAF algorithms}
Let us consider the system identification problem using adaptive filtering that at time index $n$, the desired signal $d(n)$ of the system $\bm w^o$ is given by
\begin{equation}
\label{001}
\begin{array}{rcl}
\begin{aligned}
d(n) = \bm u^\text{T}(n) \bm w^o+v(n),
\end{aligned}
\end{array}
\end{equation}
where $\bm w^o$ is an $M\times1$ sparse vector that we want to identify, $\bm u(n)=[u(n), u(n-1),...,u(n-M+1)]^\text{T}$ is the $M\times1$ input vector consisting of the recent $M$ input samples $u(n)$, and $v(n)$ is the background noise independent of $u(n)$, and $(\cdot)^\text{T}$ is a transpose operator.
\begin{figure}[htb]

    \centering
    \includegraphics[scale=0.43] {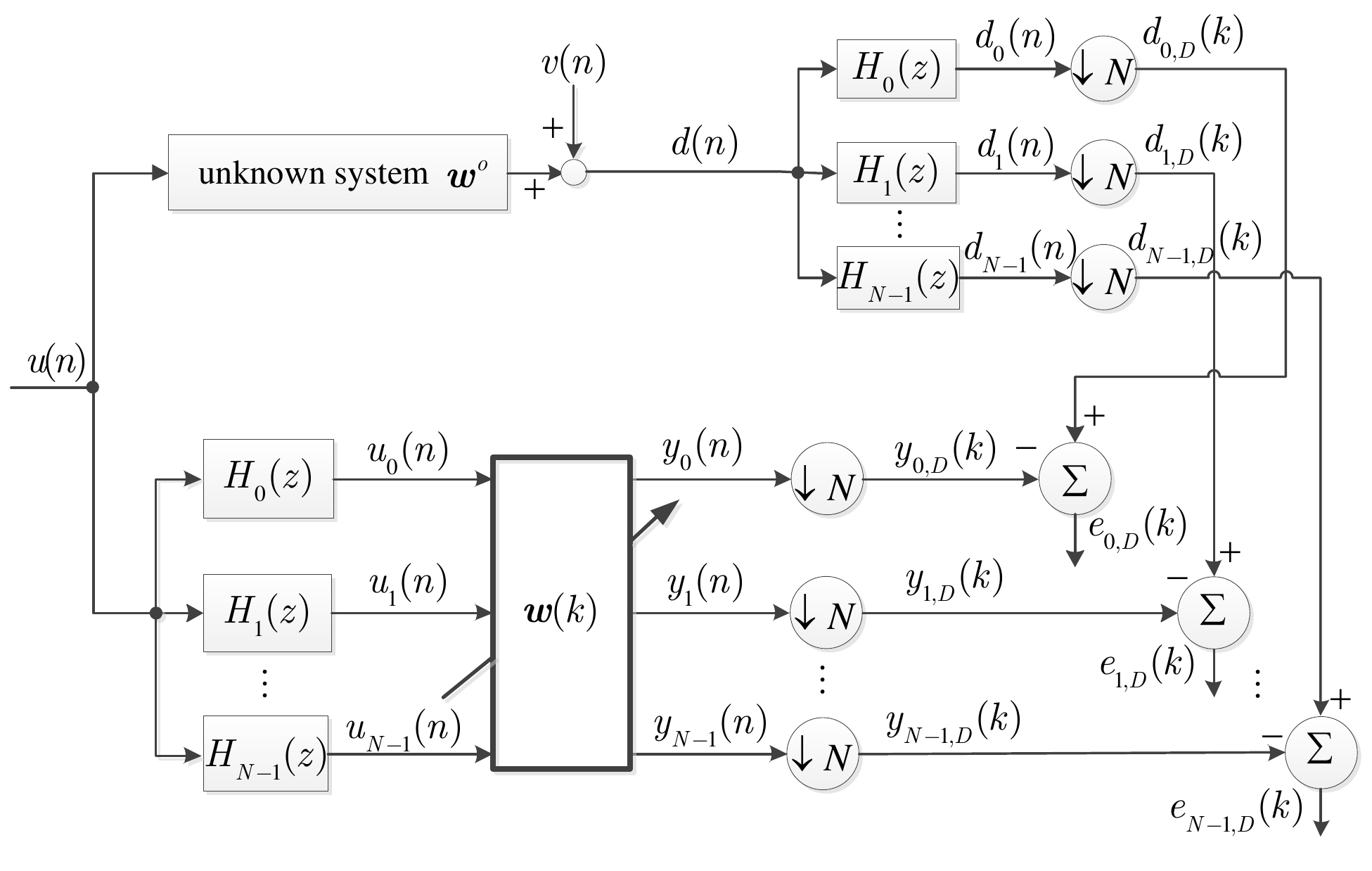}
    \vspace{-1em} \caption{Multiband-structure of SAF.}
    \label{Fig1}
\end{figure}

Fig.~\ref{Fig1} shows the multiband-structure of SAF with~$N$ subbands~\cite{lee2009subband}, where $k$ denotes the iteration index in the subband domain. By feeding $d(n)$ and $u(n)$ into the analysis filters $\{H_i(z)\}_{i=0}^{N-1}$, generate multiple subband signals $d_i(n)$ and $u_i(n)$, respectively. By filtering the subband signals $u_i(n)$ through the adaptive filter denoted by the weight vector~$\bm w(k)=[w_1(k),w_2(k),...,w_M(k)]^\text{T}$, we obtain the output signals $y_i(n)$. Then, both $d_i(n)$ and $y_i(n)$ are $N$-fold decimated to generate $d_{i,D}(k)$ and $y_{i,D}(k)$, respectively, namely, i.e., $d_{i,D}(k)=d_i(kN)$ and $y_{i,D}(k)=\bm u^\text{T}_i(k) \bm w(k)$, where $\bm u_i(k)=[u_i(kN),u_i(kN-1),...,u_i(kN-M+1)]^\text{T}$. Accordingly, the decimated error signal at each subband $i$ is formulated as
\begin{equation}
\label{002}
\begin{array}{rcl}
\begin{aligned}
e_{i,D}(k) &= d_{i,D}(k) - y_{i,D}(k)\\
&=d_{i,D}(k) - \bm u^\text{T}_i(k) \bm w(k),
\end{aligned}
\end{array}
\end{equation}
which determines how to adjust $\bm w(k)$. Thus, $\bm w(k)$ is an estimate of~$\bm w^o$ at iteration $k$. For this purpose, the NSAF algorithm is described as~\cite{lee2004improving}
\begin{equation}
\label{002x1}
\begin{array}{rcl}
\begin{aligned}
\bm w(k+1) = \bm w(k) + \mu \sum_{i=0}^{N-1} \frac{\bm u_i(k) e_{i,D}(k)}{||\bm u_{i}(k)||_2^2},
\end{aligned}
\end{array}
\end{equation}
where $\mu$ is the step-size and $||\bm x||_2$ denotes the $l_2$-norm of a vector. Considering the sparsity of $\bm w^o$, the PNSAF algorithm modifies \eqref{002x1} to
\begin{equation}
\label{002x2}
\begin{array}{rcl}
\begin{aligned}
\bm w(k+1) = \bm w(k) + \mu \sum_{i=0}^{N-1} \frac{\bm G(k) \bm u_i(k) e_{i,D}(k)}{||\bm u_{i}(k)||_{\bm G(k)}^2}.
\end{aligned}
\end{array}
\end{equation}
where the notation $||\bm x||_{\bm G} \triangleq \sqrt{\bm x^\text{T} \bm G \bm x}$ denotes the weighted $l_2$-norm of a vector. The matrix $\bm G(k)$ is diagonal with size of $M\times M$, also called the proportionate matrix, i.e., $\bm G(k)\triangleq\text{diag}\{g_1(k),...,g_M(k)\}$, and its role is to allocate an individual gain to every weight~$\{w_m(k)\}_{m=1}^M$. Different rules for calculating $\bm G(k)$ will affect the PNSAF performance~\cite{abadi2009proportionate,abadi2011family}, but this is not the focus of this paper. As such, we choose a cost-effective proportionate rule first given in~\cite{benesty2002improved}:
\begin{equation}
\label{002x3}
\begin{array}{rcl}
\begin{aligned}
g_m(k) = \frac{1-\zeta}{2M} + (1+\zeta)\frac{|w_m(k)|}{2\sum_{m=1}^M|w_m(k)| + \epsilon}
\end{aligned}
\end{array}
\end{equation}
for all $m$, where $\epsilon$ is to avoid the division by zero since $\bm w(k)$ is initialized as a null vector. In applications, typical values of~$\zeta$ are~0 or~$-0.5$. Note that, the NSAF algorithm is a special form of the PNSAF algorithm for the case of the identity matrix~$\bm G(k)$.

\section{Proposed PFBS-PNSAF Algorithm}
Consider the following minimization problem:
\begin{equation}
\label{003}
\begin{array}{rcl}
\begin{aligned}
\bm w(k+1) = \arg \min \limits_{\bm w} [J(\bm w) + \beta F(\bm w)],
\end{aligned}
\end{array}
\end{equation}
where $J(\bm w)$ is a differentiable cost function on $\bm w$ with the role of a data fitting term, and the penalty term $\beta F(\bm w)$ favors the sparsity of $\bm w$ with $\beta>0$ being the penalty intensity parameter.

Applying the PFBS technique~\cite{yamagishi2011acceleration,jeong2018automatic}, the solution to~\eqref{003} includes two steps. In the forward step, the intermediate estimate $\bm \psi(k+1)$ is obtained by solving
\begin{equation}
\label{004}
\begin{array}{rcl}
\begin{aligned}
\bm \psi(k+1) &= \arg \min \limits_{\bm w} J(\bm w),\; J(\bm w) = ||\bm w - \bm w(k)||_{\bm Q}^2 + \\
&\;\;\;\; ||\bm d(k)-\bm U^\text{T}(k)\bm w||_{(\bm U^\text{T}(k) \bm G(k) \bm U(k))^{-1}}^2,
\end{aligned}
\end{array}
\end{equation}
where $\bm d(k) \triangleq [d_{0,D}(k),d_{1,D}(k),...,d_{N-1,D}(k)]^\text{T}$, $\bm U(k) \triangleq [\bm u_{0}(k),u_{1}(k),...,\bm u_{N-1}(k)]$, and $\bm Q$ is a positive definite matrix. By setting the derivative of~\eqref{004} with respect to $\bm w$ to be zero, we acquire the following recursion
\begin{equation}
\label{004x1}
\begin{array}{rcl}
\begin{aligned}
\bm \psi(k+1) = &\bm w(k) + \left[ \bm Q + \bm U(k)(\bm U^\text{T}(k) \bm G(k) \bm U(k))^{-1}\bm U^\text{T}(k) \right]^{-1}\\
&\times \bm U(k) (\bm U^\text{T}(k) \bm G(k) \bm U(k))^{-1} \bm e_D(k),
\end{aligned}
\end{array}
\end{equation}
where $\bm e_D(k)\triangleq[e_{0,D}(k),e_{1,D}(k),...,e_{N-1,D}(k)]^\text{T}$. Exploiting the fact that $(\bm U^\text{T}(k) \bm G(k) \bm U(k))$ is an approximately diagonal matrix due to the negligible off-diagonal elements~\cite{zheng2017robust}, we are able to simplify~\eqref{004x1} as
\begin{equation}
\label{004x2}
\begin{array}{rcl}
\begin{aligned}
\bm \psi&(k+1) = \bm w(k) + \\
&\underbrace{\left[ \bm Q + \sum_{i=0}^{N-1} \frac{\bm u_i(k) \bm u_i^\text{T}(k)}{||\bm u_{i}(k)||_{\bm G(k)}^2} \right]^{-1} \sum_{i=0}^{N-1} \frac{\bm u_i(k) e_{i,D}(k)}{||\bm u_{i}(k)||_{\bm G(k)}^2}}_\text{innovation term}.
\end{aligned}
\end{array}
\end{equation}

By introducing the step-size~$\mu$ into the innovation term in~\eqref{004x2} so that flexibly controlling the algorithm performance, and then setting $\bm Q = \bm G^{-1}(k) - \sum_{i=0}^{N-1} \frac{\bm u_i(k) \bm u_i^\text{T}(k)}{||\bm u_{i}(k)||_{\bm G(k)}^2}$, the forward update formula of the proposed PFBS-PNSAF algorithm is established:
\begin{equation}
\label{005}
\begin{array}{rcl}
\begin{aligned}
\bm \psi&(k+1) = \bm w(k) + \mu \sum_{i=0}^{N-1} \frac{\bm G(k) \bm u_i(k) e_{i,D}(k)}{||\bm u_{i}(k)||_{\bm G(k)}^2 + \delta},
\end{aligned}
\end{array}
\end{equation}
where $\delta>0$ is usually a regularization parameter to prevent from~the numerical divergence when the input signal has clusters of zero values (such as the silent area of speech signal).

Subsequently, the proximal step is formulated as
\begin{equation}
\label{006}
\begin{array}{rcl}
\begin{aligned}
\bm w(k+1) &= \arg \min \limits_{\bm w}\beta F(\bm w) + \frac{1}{2\mu}||\bm \psi(k+1) - \bm w||_2^2\\
&\triangleq \text{prox}_{\mu\beta}^{F}(\bm \psi(k+1)),
\end{aligned}
\end{array}
\end{equation}
where $\text{prox}(\cdot)$ denotes the proximal operator of index~$\mu \beta$. Using the $l_1$-norm of~$\bm w$ to express the sparse penalty term~$F(\bm w)$,~\eqref{006} becomes
\begin{equation}
\label{006x1}
\begin{array}{rcl}
\begin{aligned}
\bm w(k+1) &= \arg \min \limits_{\bm w}\beta ||\bm w||_1 + \frac{1}{2\mu}||\bm \psi(k+1) - \bm w||_2^2\\
&\triangleq \text{prox}_{\mu\beta}^{l_1}(\bm \psi(k+1)).
\end{aligned}
\end{array}
\end{equation}

In the light of the optimality condition for~\eqref{006x1}, that is, zero belongs to the subgradient set at the minimizer $\text{prox}_{\mu\beta}^{l_1}(\bm \psi(k+1))$, we have
\begin{equation}
\label{006x2}
\begin{array}{rcl}
\begin{aligned}
\bm 0 \in \beta \text{sgn}(\bm w) - \frac{1}{\mu}(\bm \psi(k+1) - \bm w),
\end{aligned}
\end{array}
\end{equation}
where $\text{sgn}(\cdot)$ is the sign function. Benefited from the soft-thresholding approach~\cite{parikh2014proximal}, we obtain the closed-form solution from~\eqref{006x2}:
\begin{equation}
\label{006x3}
\begin{array}{rcl}
\begin{aligned}
\text{prox}_{\mu\beta}^{l_1}(\bm \psi(k+1)) = & \max[|\bm \psi(k+1)|-\mu\beta,\;0] \\
&\odot \text{sgn}(\bm \psi(k+1)),
\end{aligned}
\end{array}
\end{equation}
where $\odot$ denotes the element-wise product of two vectors.

\textit{Remark 1:} 
As the forward step of the PFBS-PNSAF algorithm, equation~\eqref{005} behaves like the PNSAF recursion. It is to notice from~\eqref{006x3} that the proximal step of the algorithm cuts off the components having smaller absolute values than the threshold $\mu\beta$. Intuitively, $\beta$ affects the performance of this algorithm, however, an approach to adjust it will be discussed in the following sections. Both steps promote the utilization of the underlying sparsity for the parameter vector $\bm w^o$ as much as possible, thereby enhancing the algorithm performance. On the other hand, the proposed proximal step has the same form as the shrinkage procedure in the online linearized Bregman iteration based sparse LMS algorithm~\cite{hu2014sparse,lunglmayr2016efficient,lunglmayr2017scaled}, but our derivation can be easily extended to other algorithms such as used the proportionate and normalized techniques here. Remarkably, in addition to the $l_1$-norm penalty in~\eqref{006x1}, other sparse penalties (e.g., the reweighted $l_1$-norm and the $l_0$-norm~\cite{de2014sparsity}) can also be used; as such, we can easily obtain a proximal operator different from~\eqref{006x3}. However, discussing the effect of different sparse penalties on the PFBS-PNSAF's performance is beyond the scope of this paper.

\section{Performance analysis}
In this section, we study the statistical performance of the proposed PFBS-PNSAF algorithm. As usual, the performance analysis of the proportionate algorithm is a daunting task, owing mainly to the presence of $\bm G(k)$ depending yet on $\bm w(k)$ in both the numerator and denominator of the innovation term. Moreover, the proximal step of the proposed algorithm makes its analysis further complicated. Thus, to acquire some insights on the algorithm performance, we must resort to some commonly used assumptions. To this end, the algorithm for updating the weights is rewritten as
\begin{subequations} \label{eq:21}
    \begin{align}
    \bm \psi(k+1) &= \bm w(k) + \mu \sum_{i=0}^{N-1} \frac{\bm G(k) \bm u_i(k) e_{i,D}(k)}{||\bm u_{i}(k)||_{\bm G(k)}^2+\delta}, \label{eq:21a}\\
    \bm w(k+1) &= \max[|\bm \psi(k+1)|-\mu\beta,\;0] \odot \text{sgn}(\bm \psi(k+1)).\label{eq:21b}
    \end{align}
\end{subequations}

Defining that $\bm h_i$ is the impulse response of the $i$-th analysis filter with length~$L$, then at the $i$-th subband we have
\begin{equation}
\label{022}
\begin{array}{rcl}
\begin{aligned}
\bm u_i(k) = [\bm u(kN),\bm u(kN-1),...,\bm u(kN-L+1)] \bm h_i, \\
v_{i,D}(k) = [v(kN),v(kN-1),...,v(kN-L+1)] \bm h_i,
\end{aligned}
\end{array}
\end{equation}
where $v_{i,D}(k)$ denotes the decimated subband noise by filtering the background noise $v(n)$ through the $i$-th subband analysis filter. So, the decimated subband desired signal $d_{i,D}(k)$ can be represented as
\begin{equation}
\label{023}
\begin{array}{rcl}
\begin{aligned}
d_{i,D}(k)=\bm u_i^\text{T}(k) \bm w^o + v_{i,D}(k),
\end{aligned}
\end{array}
\end{equation}
which further let~\eqref{002} become
\begin{equation}
\label{024}
\begin{array}{rcl}
\begin{aligned}
e_{i,D}(k)= e_{i,a}(k) + v_{i,D}(k),
\end{aligned}
\end{array}
\end{equation}
where $e_{i,a}(k) \triangleq \bm u^\text{T}_i(k) \widetilde{\bm w}(k)$ denotes the \emph{a priori} decimated subband error and $\widetilde{\bm w}(k)\triangleq \bm w^o - \bm w(k)$ is the weights error vector. With~\eqref{024}, we can rearrange~\eqref{eq:21a} as
\begin{equation}
\label{025}
\begin{array}{rcl}
\begin{aligned}
\widetilde{\bm \psi}(k+1) = & \widetilde{\bm w}(k) - \mu \sum_{i=0}^{N-1} \frac{\bm G(k) \bm u_i(k) e_{i,a}(k)}{||\bm u_{i}(k)||_{\bm G(k)}^2+\delta}  \\
 &-\mu \sum_{i=0}^{N-1} \frac{\bm G(k) \bm u_i(k) v_{i,D}(k)}{||\bm u_{i}(k)||_{\bm G(k)}^2+\delta},
\end{aligned}
\end{array}
\end{equation}
where $\widetilde{\bm \psi}(k)\triangleq \bm w^o - \bm \psi(k)$. By introducing the auxiliary function $P_{\mu \beta}(\cdot)$,
\begin{equation}
\label{026}
P_{\mu \beta}(x) = \left\{ \begin{aligned}
&\mu \beta\text{sgn}(x), \text{ if } |x| > \mu\beta\\
&x, \text{ if } |x| \leq \mu\beta
\end{aligned} \right.
\end{equation}
so that $P_{\mu \beta}(\bm \psi(k+1)) = [P_{\mu \beta}(\psi_1(k+1)),P_{\mu \beta}(\psi_2(k+1)),...,P_{\mu \beta}(\psi_M(k+1))]^\text{T}$, we formulate~\eqref{eq:21b} as
\begin{equation}
\label{027x1}
\begin{array}{rcl}
\begin{aligned}
\bm w(k+1) = \bm \psi(k+1) - P_{\mu \beta}(\bm \psi(k+1)).
\end{aligned}
\end{array}
\end{equation}
Furthermore, by subtracting~\eqref{027x1} from $\bm w^o$, we obtain
\begin{equation}
\label{027x2}
\begin{array}{rcl}
\begin{aligned}
\widetilde{\bm w}(k+1) = \widetilde{\bm \psi}(k+1) + P_{\mu \beta}(\bm \psi(k+1)).
\end{aligned}
\end{array}
\end{equation}

Combining~\eqref{025} with \eqref{027x2}, the recursion of $\widetilde{\bm w}(k)$ is established:
\begin{subequations} \label{eq:28}
    \begin{align}
    \begin{array}{rcl}
    \begin{aligned}
    \widetilde{\bm \psi}(k+1) =& \widetilde{\bm w}(k) - \mu \sum_{i=0}^{N-1} \frac{\bm G(k) \bm u_i(k) e_{i,a}(k)}{||\bm u_{i}(k)||_{\bm G(k)}^2+\delta} \\
    &- \mu \sum_{i=0}^{N-1} \frac{\bm G(k) \bm u_i(k) v_{i,D}(k)}{||\bm u_{i}(k)||_{\bm G(k)}^2+\delta},\end{aligned}
    \end{array} \label{eq:28a}\\
    \widetilde{\bm w}(k+1) = \widetilde{\bm \psi}(k+1) + P_{\mu \beta}(\bm \psi(k+1)).\label{eq:28b}
    \end{align}
\end{subequations}

To continue the analysis, the following assumptions are made.

\textbf{Assumption 1}: The input signal $u(n)$ is wide-sense stationary random process with zero-mean and positive definite autocorrelation matrix $\bm R = \text{E}\{\bm u(n)\bm u^\text{T}(n)\}$.

\textbf{Assumption 2}: The background noise $v(n)$ is zero mean white random process with variance $\sigma_v^2$.

\textbf{Assumption 3}: The weight error vector $\widetilde{\bm w}(k)$ is statistically independent of the decimated input vectors $\{\bm u_i(k)\}_{i=0}^{N-1}$.

\textbf{Assumption 4}: $\bm G(k)$ depends on $\bm w(k)$ as evidenced in~\eqref{002x3}, but $\text{Tr}\{\bm G(k)\}=1$ which makes it vary slowly from the iteration~$k$ to the next iteration~$k+1$ as compared to $\bm w(k)$. Thus, we can assume that $\bm G(k)$ is independent of $\bm u(n)$ and $\widetilde{\bm w}(k)$ especially when near convergence.

Assumptions 1 to 3 are customary in analyzing adaptive filtering algorithms, where assumption~3 is the known~\emph{independence assumption}~\cite{sayed2003fundamentals,chen2014steady,dang2019kernel}. From \eqref{022} and assumption~1, the $i$-th subband decimated input vector has also zero-mean and positive definite autocorrelation matrix $\bm R_i = \text{E}\{\bm u_i(k)\bm u_i^\text{T}(k)\}$. From \eqref{022} and assumption~2, the decimated subband noise $\{v_{i,D}(k)\}_{i=0}^{N-1}$ are zero mean white processes with variances $\sigma_{v,i}^2=||\bm h_i||_2^2 \sigma_v^2$. Note that, the paraunitary assumption of the analysis filters leads to $||\bm h_i||_2^2=1/N$, which was frequently used in the performance analysis of the NSAF algorithm~\cite{yin2010stochastic,jeong2015mean}; however, we do not consider the paraunitary property. assumption~4 is strong especially in the transient stage, but it has been employed to simplify the analysis of the proportionate NLMS algorithm and the analytical results were also verified by simulations in~\cite{loganathan2010performance,das2016improving,haddad2014transient}.

\vspace{-0.25cm}
\subsection{Mean behavior}
Under assumptions 1 and 2, enforcing the expectations for both sides of~\eqref{eq:28} yields
\begin{subequations} \label{eq:29}
    \begin{align}
    \text{E}\{\widetilde{\bm \psi}(k+1)\} &= \text{E}\{\widetilde{\bm w}(k)\} - \mu \sum_{i=0}^{N-1} \text{E}\left\lbrace \frac{\bm G(k) \bm u_i(k) e_{i,a}(k)}{||\bm u_{i}(k)||_{\bm G(k)}^2+\delta} \right\rbrace, \label{eq:29a}\\
    \text{E}\{\widetilde{\bm w}(k+1)\} &= \text{E}\{\widetilde{\bm \psi}(k+1)\} + \text{E}\{P_{\mu \beta}(\bm \psi(k+1))\}.\label{eq:29b}
    \end{align}
\end{subequations}

For long adaptive filters, i.e., $M\gg1$, the following approximation can be made
\begin{equation}
\label{030}
\begin{array}{rcl}
\begin{aligned}
\text{E}\left\lbrace \frac{\bm G(k) \bm u_i(k) e_{i,a}(k)}{||\bm u_{i}(k)||_{\bm G(k)}^2 + \delta} \right\rbrace &\approx \frac{\text{E}\{\bm G(k) \bm u_i(k) e_{i,a}(k)\}}{\text{E}\{||\bm u_{i}(k)||_{\bm G(k)}^2\} + \delta} \\
&\stackrel{(a)}{\approx} \frac{\text{E}\{\bm G(k) \bm u_i(k) e_{i,a}(k)\}}{\sigma_{u,i}^2 + \delta}
\end{aligned}
\end{array}
\end{equation}
where the approximation $(a)$ is based on $\text{Tr}\{\bar{\bm G}(k)\}=1$ and the assumption of large enough number of subbands. Under this assumption, each of decimated subband input signals $u_i(k)$ is approximately white with variance $\sigma_{u,i}^2$, which is very efficient in designing and analyzing SAF algorithms~\cite{jeong2015mean,zhang2018mean}.

Based on assumptions 3 and 4, the approximation above lets~\eqref{eq:29a} become
\begin{equation}
\label{031}
\begin{array}{rcl}
\begin{aligned}
\text{E}\{\widetilde{\bm \psi}(k+1)\} = \left( \bm I_M - \mu \sum_{i=0}^{N-1} \frac{\bar{\bm G}(k)\bm R_i}{\sigma_{u,i}^2 + \delta} \right) \text{E}\{\widetilde{\bm w}(k)\}.
\end{aligned}
\end{array}
\end{equation}
where $\bar{\bm G}(k)\triangleq \text{E}\{\bm G(k)\}$.

Since entries of $P_{\mu \beta}(\bm \psi(k+1))$ are bounded, the convergence condition of the PFBS-PNSAF algorithm in the mean reduces to that of the recursion~\eqref{031}. Hence, $\left( \bm I_M - \mu \sum_{i=0}^{N-1} \frac{\bar{\bm G}(k)\bm R_i}{\sigma_{u,i}^2 + \delta} \right)$ is required to be a stable matrix, which leads to the following theorem:
\newtheorem{theorem}{Theorem}
\begin{theorem}
    \textit{
    The PFBS-PNSAF algorithm is convergent in the mean if, and only if the step-size satisfies
    \vspace{-0.1cm}
    \begin{equation}
    \label{032}
    \begin{array}{rcl}
    \begin{aligned}
    0 < \mu< \frac{2}{ \sum_{i=0}^{N-1} \frac{\lambda_\text{max}\left( \bar{\bm G}(k)\bm R_i \right)}{\sigma_{u,i}^2 + \delta} },
    \end{aligned}
    \end{array}
    \end{equation}
    where $\lambda_\text{max}(\cdot)$ indicates the maximum eigenvalue of a matrix.}\end{theorem}

When the algorithm has converged to the steady-state, we can obtain the following relation from~\eqref{026},~\eqref{eq:29b}, and~\eqref{031}:
    \begin{equation}
    \label{033}
    \begin{array}{rcl}
    \begin{aligned}
    \mathrm{E}\{\widetilde{\bm w}(\infty)\} = \left\{ \begin{aligned}
    &\beta \left(\sum_{i=0}^{N-1} \frac{\bar{\bm G}(\infty)\bm R_i}{\sigma_{u,i}^2 + \delta} \right)^{-1} \times \\
    &\;\;\mathrm{E}\{\mathrm{sgn}(\bm w(\infty))\}, \mathrm{ if }\;|w_m(\infty)| > \mu\beta\\
    &\bm w^o, \mathrm{ if }\; |w_m(\infty)| \leq \mu\beta,
    \end{aligned} \right.
    \end{aligned}
    \end{array}
    \end{equation}
    where $m=1,...,M$, which further becomes
 \begin{equation}
 \label{034x}
 \begin{array}{rcl}
 \begin{aligned}
 \mathrm{E}\{\bm w(\infty)\} = \left\{ \begin{aligned}
 &\bm w^o - \beta \left(\sum_{i=0}^{N-1} \frac{\bar{\bm G}(\infty)\bm R_i}{\sigma_{u,i}^2 + \delta} \right)^{-1} \times \\
 &\;\;\mathrm{E}\{\mathrm{sgn}(\bm w(\infty))\}, \mathrm{ if }\;|w_m(\infty)| > \mu\beta\\
 &\bm 0, \mathrm{ if }\; |w_m(\infty)| \leq \mu\beta.
 \end{aligned} \right.
 \end{aligned}
 \end{array}
 \end{equation}
Equation~\eqref{034x} shows that the PFBS-PNSAF algorithm always drives the filter's weights with smaller magnitude than~$\mu\beta$ to zero. This is very useful in estimation of a sparse vector~$\bm w^o$, since the majority of coefficients are zero. For these zero coefficients, the proposed PFBS-PNSAF algorithm is unbiased, correspondingly enhancing the estimation performance of those coefficients. Note that~\eqref{034x} also shows, for identifying nonzero coefficients with $|w_m^o| > \mu\beta$, the proposed algorithm is biased, but these deviations are very small as compared to the $|w_m^o|$ as $\beta$ is very small in simulations.
\vspace{-0.3cm}
\subsection{Mean square behavior}
To analyze the mean square behavior, the autocorrelation matrices of $\widetilde{\bm w}(k)$ and $\widetilde{\bm \psi}(k)$ are defined as $\widetilde{\bm W}(k) \triangleq \text{E}\{\widetilde{\bm w}(k)\widetilde{\bm w}^\text{T}(k)\}$ and $\widetilde{\bm \Xi}(k) \triangleq \text{E}\{\widetilde{\bm \psi}(k) \widetilde{\bm \psi}^\text{T}(k)\}$, respectively. Then, both sides of~\eqref{eq:28a} are multiplied by their transposes, and by taking the expectation of the equation we obtain
\begin{equation}
\label{035}
\begin{array}{rcl}
\begin{aligned}
&\widetilde{\bm \Xi}(k+1) = \widetilde{\bm W}(k) - \mu \text{E}\left\lbrace \sum_{i=0}^{N-1} \frac{\bm G(k) \bm u_i(k) e_{i,a}(k)}{||\bm u_{i}(k)||_{\bm G(k)}^2+ \delta} \widetilde{\bm w}^\text{T}(k) \right\rbrace \\
&- \mu \text{E}\left\lbrace \left( \sum_{i=0}^{N-1} \frac{\bm G(k) \bm u_i(k) e_{i,a}(k)}{||\bm u_{i}(k)||_{\bm G(k)}^2+ \delta} \widetilde{\bm w}^\text{T}(k) \right)^\text{T} \right\rbrace  \\

&+ \mu^2 \text{E}\left\lbrace \sum_{i=0}^{N-1} \frac{\bm G(k) \bm u_i(k) e_{i,a}(k)}{||\bm u_{i}(k)||_{\bm G(k)}^2+ \delta} \left( \sum_{i=0}^{N-1} \frac{\bm G(k) \bm u_i(k) e_{i,a}(k)}{||\bm u_{i}(k)||_{\bm G(k)}^2+ \delta}\right)^\text{T} \right\rbrace  \\
&+ \mu^2 \text{E}\left\lbrace \sum_{i=0}^{N-1} \frac{\bm G(k) \bm u_i(k) v_{i,D}(k)}{||\bm u_{i}(k)||_{\bm G(k)}^2+ \delta} \left( \sum_{i=0}^{N-1} \frac{\bm G(k) \bm u_i(k) v_{i,D}(k)}{||\bm u_{i}(k)||_{\bm G(k)}^2+ \delta}\right)^\text{T} \right\rbrace \\
& + \text{E} \{ \bm C_\text{cross} \}.
\end{aligned}
\end{array}
\end{equation}
where $\bm C_\text{cross}$ sums the cross terms associated with $v_{i,D}(k)$ so that its mean is zero. Using assumptions~3 and~4 and referring to~\eqref{030}, the relation~\eqref{035} further becomes
\begin{equation}
\label{036}
\begin{array}{rcl}
\begin{aligned}
&\widetilde{\bm \Xi}(k+1) = \widetilde{\bm W}(k) - \mu \bar{\bm G}(k) \sum_{i=0}^{N-1} \frac{\bm R_i}{\sigma_{u,i}^2 + \delta} \widetilde{\bm W}(k) \\
&- \mu \widetilde{\bm W}^\text{T}(k) \left( \bar{\bm G}(k) \sum_{i=0}^{N-1} \frac{\bm R_i}{\sigma_{u,i}^2 + \delta} \right)^\text{T} \\
&+ \mu^2 \bar{\bm G}(k) \underbrace{ \text{E}\left\lbrace \sum_{i=0}^{N-1} \frac{\bm u_i(k) \bm u_i^\text{T}(k) }{\sigma_{u,i}^2 + \delta} \widetilde{\bm W}(k) \sum_{j=0}^{N-1} \frac{\bm u_j(k) \bm u_j^\text{T}(k)}{\sigma_{u,j}^2 + \delta}\right\rbrace }\limits_{(b)} \bar{\bm G}(k)  \\
&+ \mu^2 \bar{\bm G}(k) \sum_{i=0}^{N-1} \frac{\sigma_{v,i}^2 \bm R_i}{(\sigma_{u,i}^2 + \delta)^2} \bar{\bm G}(k).
\end{aligned}
\end{array}
\end{equation}
where the term~$(b)$ is further calculated in~Appendix~A.

By plugging~\eqref{0A3} into~\eqref{036}, the evolution of $\widetilde{\bm W}(k)$ with iterations is established:
\begin{equation}
\label{038}
\begin{array}{rcl}
\begin{aligned}
&\widetilde{\bm \Xi}(k+1) = \widetilde{\bm W}(k) - \mu \bar{\bm G}(k) \sum_{i=0}^{N-1} \frac{\bm R_i}{\sigma_{u,i}^2 + \delta} \widetilde{\bm W}(k) \\
&- \mu \widetilde{\bm W}^\text{T}(k) \left( \bar{\bm G}(k) \sum_{i=0}^{N-1} \frac{\bm R_i}{\sigma_{u,i}^2 + \delta} \right)^\text{T} \\
&+ \mu^2 \bar{\bm G}(k) \sum_{i=0}^{N-1} \frac{1}{(\sigma_{u,i}^2 + \delta)^2} \bm R_i \widetilde{\bm W}(k) \bm R_i \bar{\bm G}(k) \\
&+ \mu^2 \bar{\bm G}(k) \sum_{i=0}^{N-1} \frac{1}{(\sigma_{u,i}^2 + \delta)^2} \bm R_i \text{Tr}\left( \widetilde{\bm W}(k) \bm R_i \right) \bar{\bm G}(k) \\
&+ \mu^2 \bar{\bm G}(k) \sum_{i=0}^{N-1} \frac{\sigma_{v,i}^2 \bm R_i}{(\sigma_{u,i}^2 + \delta)^2} \bar{\bm G}(k).
\end{aligned}
\end{array}
\end{equation}

By taking the autocorrelation matrices for both sides of~\eqref{eq:28b}, we obtain
\begin{equation}
\label{039}
\begin{array}{rcl}
\begin{aligned}
\widetilde{\bm W}(k+1) =& \widetilde{\bm \Xi}(k+1) + \bm \Psi(k+1) \\
&+ \bm \Psi^\text{T}(k+1) + \bm \Theta(k+1),
\end{aligned}
\end{array}
\end{equation}
where
\begin{subequations} \label{eq:40}
    \begin{align}
    \bm \Psi(k+1) &\triangleq \text{E}\{ \widetilde{\bm \psi}(k+1) P_{\mu \beta}^\text{T}(\bm \psi(k+1))\}, \label{eq:40b} \\
    \bm \Theta(k+1) &\triangleq \text{E}\left\lbrace P_{\mu \beta}(\bm \psi(k+1)) P_{\mu \beta}^\text{T}(\bm \psi(k+1)) \right\rbrace.\label{eq:41b}
    \end{align}
\end{subequations}

According to the definitions of the MSD and excess mean square error (EMSE) respectively for the algorithm~\cite{lee2009subband}, i.e.,
\begin{equation}
\label{041}
\begin{array}{rcl}
\begin{aligned}
\text{MSD}(k) &\triangleq \text{E}\{||\widetilde{\bm w}(k)||_2^2\} = \text{Tr}\{\widetilde{\bm W}(k)\},\\
\text{EMSE}(k) &\triangleq \frac{1}{N} \sum_{i=0}^{N-1} \text{E}\{e_{i,a}^2(k)\} = \text{Tr}\{\widetilde{\bm W}(k) \bm R_i\},
\end{aligned}
\end{array}
\end{equation}
it follows that~\eqref{038} and~\eqref{039} model the mean square evolution behavior of the PFBS-PNSAF algorithm. To implement~\eqref{038} and~\eqref{039}, the remaining problem is how to compute the moments~$\bar{\bm G}(k)$, $\text{E}\{P_{\mu \beta}(\bm \psi(k+1)) \}$, $\bm \Psi(k+1)$, and $\bm \Theta(k+1)$ at the iteration~$k$. For this purpose, we employ the element-wise approach, i.e., $\bar{g}_m(k)=\text{E}\{g_m(k)\}$, $\bm \Psi_{m,l}(k+1) = \text{E}\{ \widetilde{w}_m(k+1) P_{\mu \beta}(w_l(k+1)) \}$, and $\bm \Theta_{m,l}(k+1) = \text{E}\left\lbrace P_{\mu \beta}(w_m(k+1)) P_{\mu \beta}(w_l(k+1)) \right\rbrace$ for $m,l=1,...,M$. In addition, two additional assumptions are considered:

\textbf{Assumption 5}: The $m$-th component of the weight error vector $\widetilde{\bm w}(k)$ at iteration $k$, has a Gaussian distribution, namely, $\widetilde{w}_m(k)\sim \aleph(z_{w,m}(k),\sigma_{w,m}^2(k))$, where the mean $z_{w,m}(k)$ is the $m$-th component of $\text{E}\{\widetilde{\bm w}(k)\}$ from~\eqref{eq:29b} and the variance $\sigma_{w,m}^2(k)$ is computed by $\sigma_{w,m}^2(k) = \widetilde{\bm W}_{m,m}(k)-z_{w,m}^2(k)$. Similarly, for $\widetilde{\psi}_m(k+1)$ we have $\widetilde{\psi}_m(k+1) \sim \aleph(z_{\psi,m}(k+1),\sigma_{\psi,m}^2(k+1))$, where $z_{\psi,m}(k+1)$ is the $m$-th component of $\text{E}\{\widetilde{\bm \psi}(k+1)\}$ from~\eqref{031} and $\sigma_{\psi,m}^2(k+1) = \widetilde{\bm \Xi}_{m,m}(k+1)-z_{\psi,m}^2(k+1)$.

\textbf{Assumption 6}: If $m\neq l$, we assume $\text{E}\{\psi_m(k+1) P_{\mu \beta}(\psi_l(k+1)) \}\approx \text{E}\{\psi_m(k+1)\} \text{E}\{P_{\mu \beta}(\psi_l(k+1)) \}$ and $\text{E}\{ P_{\mu \beta}(\psi_m(k+1)) P_{\mu \beta}(\psi_l(k+1)) \} \approx \text{E}\{ P_{\mu \beta}(\psi_m(k+1)) \} \text{E}\{ P_{\mu \beta}(\psi_l(k+1)) \}$.

These two assumptions have been used in the literature on the analysis of sparse adaptive filtering algorithms~\cite{haddad2014transient,yu2019sparsity,loganathan2010performance,haddad2014transient}. Assumption~5 can be supported by the central limit theorem. The separable assumption~6 is strong, but it leads to the simplification of the analysis.

\textit{1) Calculation of $\bar{g}_m(k)$:} From~\eqref{002x3}, we have~\cite{loganathan2010performance,haddad2014transient}
\begin{equation}
\label{042}
\begin{array}{rcl}
\begin{aligned}
\bar{g}_m(k) = \frac{1-\zeta}{2M} + (1+\zeta)\frac{\text{E}\{|w_m(k)|\}}{2\sum_{m=1}^M\text{E}\{|w_m(k)|\} + \epsilon}.
\end{aligned}
\end{array}
\end{equation}

Based on assumption~5, $w_m(k)$ follows the distribution~$\aleph(\bar{x},\sigma_x^2)$ with $\bar{x}=w_m^o-z_m(k)$ and $\sigma_x^2=\sigma_m^2(k)$ where $w_m^o$ is the $m$-th component of $\bm w^o$, therefore,
\begin{equation}
\label{043}
\begin{array}{rcl}
\begin{aligned}
\text{E}\{|x|\} =& \frac{1}{\sqrt{2\pi}\sigma_x} \int_{-\infty}^{\infty}|x|\exp^{-\left( \frac{x-\bar{x}}{\sqrt{2}\sigma_x}\right)^2}dx \\
=&\sqrt{\frac{2}{\pi}} \sigma_x \exp^{-\frac{\bar{x}^2}{2\sigma_x^2}} + \bar{x} \text{erf}\left( \frac{\bar{x}}{\sqrt{2}\sigma_x}\right)
\end{aligned}
\end{array}
\end{equation}
where $\text{erf}(x)=\frac{2}{\sqrt{\pi}}\int_{0}^{x}\exp^{-t^2}dt$.

\textit{2) Calculation of $\bm \Psi_{m,l}(k+1)$:} It is rewritten as $\bm \Psi_{m,l}(k+1) = w_m^o\text{E}\{P_{\mu \beta}(\psi_l(k+1)) \} - \text{E}\{\psi_m(k+1) P_{\mu \beta}(\psi_l(k+1)) \}$. Likewise, since $\psi_m(k+1)$ follows the distribution~$\aleph(\bar{x},\sigma_x^2)$ but with $\bar{x}=w_m^o-z_{\psi,m}(k+1)$ and $\sigma_x^2=\sigma_{\psi,m}^2(k+1)$, $\text{E}\{P_{\mu \beta}(\psi_l(k+1)) \}$ is computed by~\eqref{044}, where $a_1 = \frac{\mu\beta + \bar{x}}{\sqrt{2}\sigma_x}$ and $a_2 = \frac{\mu\beta - \bar{x}}{\sqrt{2}\sigma_x}$. When $m=l$, we compute $\text{E}\{\psi_m(k+1) P_{\mu \beta}(\psi_m(k+1)) \}$ by~\eqref{045}.

\textit{3) Calculation of $\bm \Theta_{m,l}(k+1)$:} When $m=l$, $\bm \Theta_{m,m}(k+1)=\text{E}\{ P_{\mu \beta}^2(\psi_m(k+1))\}$ is obtained by~\eqref{046}.

\newcounter{mytempeqncnt}
\begin{figure*}[!t]
    \normalsize
    \setcounter{mytempeqncnt}{\value{equation}}
    \setcounter{equation}{37}
    \begin{equation}
    \label{044}
    \begin{array}{rcl}
    \begin{aligned}
    \text{E}\{P_{\mu \beta}(x)\} &= \frac{1}{\sqrt{2\pi}\sigma_x} \int_{-\infty}^{\infty} P_{\mu \beta}(x) \exp^{-\left( \frac{x-\bar{x}}{\sqrt{2}\sigma_x}\right)^2}dx \\
    \stackrel{(20)}{=}& - \frac{\mu\beta}{\sqrt{2\pi}\sigma_x} \int_{-\infty}^{-\mu\beta} \exp^{-\left( \frac{x-\bar{x}}{\sqrt{2}\sigma_x}\right)^2}dx + \frac{1}{\sqrt{2\pi}\sigma_x} \int_{-\mu\beta}^{\mu\beta} x \exp^{-\left( \frac{x-\bar{x}}{\sqrt{2}\sigma_x}\right)^2}dx +  \frac{\mu\beta}{\sqrt{2\pi}\sigma_x} \int_{\mu\beta}^{\infty} \exp^{-\left( \frac{x-\bar{x}}{\sqrt{2}\sigma_x}\right)^2}dx \\
    =&\frac{1}{2}\sqrt{\frac{2}{\pi}} \sigma_x (\exp^{-a_1^2}-\exp^{-a_2^2}) + \left( \frac{\bar{x}}{2} + \frac{\mu\beta}{2} \right) \text{erf}(a_1) +  \left( \frac{\bar{x}}{2} - \frac{\mu\beta}{2} \right) \text{erf}(a_2).
    \end{aligned}
    \end{array}
    \end{equation}
    \begin{equation}
    \label{045}
    \begin{array}{rcl}
    \begin{aligned}
    \text{E}&\{xP_{\mu \beta}(x)\} = \frac{1}{\sqrt{2\pi}\sigma_x} \int_{-\infty}^{\infty} xP_{\mu \beta}(x) \exp^{-\left( \frac{x-\bar{x}}{\sqrt{2}\sigma_x}\right)^2}dx \\
    &\stackrel{(20)}{=} - \frac{\mu\beta}{\sqrt{2\pi}\sigma_x} \int_{-\infty}^{-\mu\beta} x\exp^{-\left( \frac{x-\bar{x}}{\sqrt{2}\sigma_x}\right)^2}dx + \frac{1}{\sqrt{2\pi}\sigma_x} \int_{-\mu\beta}^{\mu\beta} x^2 \exp^{-\left( \frac{x-\bar{x}}{\sqrt{2}\sigma_x}\right)^2}dx +  \frac{\mu\beta}{\sqrt{2\pi}\sigma_x} \int_{\mu\beta}^{\infty} x\exp^{-\left( \frac{x-\bar{x}}{\sqrt{2}\sigma_x}\right)^2}dx \\
    &= \left( \frac{\mu \beta}{2}\sqrt{\frac{2}{\pi}} \sigma_x + \sqrt{\frac{2}{\pi}} \sigma_x \bar{x} - \frac{\sigma_x^2}{\sqrt{\pi}}a_1 \right) \exp^{-a_1^2}  + \left( \frac{\mu \beta}{2}\sqrt{\frac{2}{\pi}} \sigma_x - \sqrt{\frac{2}{\pi}} \sigma_x \bar{x} - \frac{\sigma_x^2}{\sqrt{\pi}}a_2 \right) \exp^{-a_2^2} \\
    &+\frac{1}{2}(\sigma_x^2 + \bar{x}^2 + \mu \beta \bar{x}) \text{erf}(a_1)
    + \frac{1}{2}(\sigma_x^2 + \bar{x}^2 - \mu \beta \bar{x}) \text{erf}(a_2).
    \end{aligned}
    \end{array}
    \end{equation}
    \begin{equation}
    \label{046}
    \begin{array}{rcl}
    \begin{aligned}
    \text{E}&\{P_{\mu \beta}^2(x)\} = \frac{1}{\sqrt{2\pi}\sigma_x} \int_{-\infty}^{\infty} P_{\mu \beta}^2(x) \exp^{-\left( \frac{x-\bar{x}}{\sqrt{2}\sigma_x}\right)^2}dx \\
    &\stackrel{(20)}{=} \frac{\mu^2\beta^2}{\sqrt{2\pi}\sigma_x} \int_{-\infty}^{-\mu\beta} \exp^{-\left( \frac{x-\bar{x}}{\sqrt{2}\sigma_x}\right)^2}dx + \frac{1}{\sqrt{2\pi}\sigma_x} \int_{-\mu\beta}^{\mu\beta} x^2 \exp^{-\left( \frac{x-\bar{x}}{\sqrt{2}\sigma_x}\right)^2}dx + \frac{\mu^2\beta^2}{\sqrt{2\pi}\sigma_x} \int_{\mu\beta}^{\infty} x\exp^{-\left( \frac{x-\bar{x}}{\sqrt{2}\sigma_x}\right)^2}dx \\
    &= \left(\sqrt{\frac{2}{\pi}} \sigma_x \bar{x} - \frac{\sigma_x^2}{\sqrt{\pi}}a_1 \right) \exp^{-a_1^2} - \left(\sqrt{\frac{2}{\pi}} \sigma_x \bar{x} + \frac{\sigma_x^2}{\sqrt{\pi}}a_2 \right) \exp^{-a_2^2} +\frac{1}{2}(\sigma_x^2 + \bar{x}^2 - \mu^2 \beta^2) (\text{erf}(a_1) + \text{erf}(a_2)) + \mu^2 \beta^2.
    \end{aligned}
    \end{array}
    \end{equation}
    \setcounter{equation}{\value{mytempeqncnt}}
    \hrulefill
    \vspace*{4pt}
\end{figure*}
\addtocounter{equation}{3}

Note that, when $m\neq l$, $\bm \Psi_{m,l}(k+1)$ and $\bm \Theta_{m,l}(k+1)$ can be obtained from assumption~6.

Obviously, when the recursions~\eqref{038} and~\eqref{039} reach the steady-state, we can obtain the steady-state MSD or EMSE. However, from the above recursions we can not obtain some intuitive insights on the steady-state performance, due mainly to the existence of $\bar{\bm G}(k)$. Although we can try to impose the vectorization operation and the Kronecker product~\cite{sayed2003fundamentals} on~\eqref{038} and~\eqref{039} to solve this problem, this brings about the inverse matrix with the size~$M^2\times M^2$ which is not suitable for the case of large~$M$. In the sequel, therefore we show the steady-state behavior and mean-square convergence condition for the algorithm from an alternative approach. By performing the squared weighted $l_2$-norm for both sides of~\eqref{eq:28a} and ~\eqref{eq:28b} respectively with the weighted matrix $\bm G^{-1}(k)$ and then taking their expectations over assumption~2, we find the following relations
\begin{equation}
\label{046x1}
\begin{array}{rcl}
\begin{aligned}
&\text{E}\left\lbrace ||\widetilde{\bm \psi}(k+1)||_{\bm G^{-1}(k)}^2\right\rbrace  = \text{E} \left\lbrace ||\widetilde{\bm w}(k)||_{\bm G^{-1}(k)}^2\right\rbrace-   \\
&2\mu\text{E} \left\lbrace \sum_{i=0}^{N-1} \frac{e_{i,a}^2(k)}{||\bm u_{i}(k)||_{\bm G(k)}^2+\delta} \right\rbrace + \\
& \mu^2 \text{E}\left\lbrace \sum_{i=0}^{N-1} \frac{||\bm u_{i}(k)||_{\bm G(k)}^2 e_{i,a}^2(k)}{(||\bm u_{i}(k)||_{\bm G(k)}^2+\delta)^2} \right\rbrace +\\
& \mu^2 \text{E}\left\lbrace \sum_{i=0}^{N-1} \sum_{j=0,j\neq i}^{N-1} \frac{\overbrace{\bm u_i^\text{T}(k) \bm G(k) \bm u_j(k)}\limits^{(c)} e_{i,a}(k)e_{j,a}(k)}{(||\bm u_{i}(k)||_{\bm G(k)}^2+\delta) (||\bm u_{j}(k)||_{\bm G(k)}^2+\delta)} \right\rbrace + \\
& \mu^2 \text{E}\sum_{i=0}^{N-1} \left\lbrace \frac{||\bm u_{i}(k)||_{\bm G(k)}^2}{(||\bm u_{i}(k)||_{\bm G(k)}^2+\delta)^2} \right\rbrace ||\bm h_i||_2^2\sigma_v^2, \\
\end{aligned}
\end{array}
\end{equation}
and
\begin{equation}
\label{046x2}
\begin{array}{rcl}
\begin{aligned}
\text{E} &\left\lbrace ||\widetilde{\bm w}(k+1)||_{\bm G^{-1}(k)}^2\right\rbrace =  \text{E}\left\lbrace ||\widetilde{\bm \psi}(k+1)||_{\bm G^{-1}(k)}^2\right\rbrace \\
&\;\;\;\;+ 2 \text{E}\left\lbrace \widetilde{\bm \psi}^\text{T}(k+1) \bm G^{-1}(k) P_{\mu \beta}(\bm \psi(k+1))\right\rbrace \\
&\;\;\;\;+ \text{E} \left\lbrace P_{\mu \beta}^\text{T}(\bm \psi(k+1)) \bm G^{-1}(k) P_{\mu \beta}(\bm \psi(k+1))\right\rbrace.
\end{aligned}
\end{array}
\end{equation}

Recalling again the term $(c)$ in~\eqref{046x1} approximates zero when $j\neq i$~\cite{zheng2017robust}, and applying assumptions 3 and 4 for a long adaptive filter, we further simplify~\eqref{046x1} as
\begin{equation}
\label{046x3}
\begin{array}{rcl}
\begin{aligned}
\text{E}&\left\lbrace ||\widetilde{\bm w}(k+1)||_{\bm G^{-1}(k)}^2\right\rbrace  = \text{E} \left\lbrace ||\widetilde{\bm w}(k)||_{\bm G^{-1}(k)}^2\right\rbrace  \\
&- 2 \mu \sum_{i=0}^{N-1} \frac{\text{E}\{e_{i,a}^2(k)\}}{\sigma_{u,i}^2+\delta} + \mu^2 \sum_{i=0}^{N-1} \frac{\sigma_{u,i}^2 \text{E}\{e_{i,a}^2(k)\}}{(\sigma_{u,i}^2+\delta)^2}  \\
&+ \mu^2 \sum_{i=0}^{N-1} \frac{\sigma_{u,i}^2 ||\bm h_i||_2^2\sigma_v^2}{(\sigma_{u,i}^2+\delta)^2}.
\end{aligned}
\end{array}
\end{equation}

Since $\text{E}\left\lbrace ||\widetilde{\bm w}(k+1)||_{\bm G^{-1}(k)}^2\right\rbrace  = \text{E} \left\lbrace ||\widetilde{\bm w}(k)||_{\bm G^{-1}(k)}^2\right\rbrace $ in the steady-state, we impose the limits on both sides of~\eqref{046x2} and~\eqref{046x3} as $k \rightarrow \infty$, yielding
\begin{equation}
\label{047}
\begin{array}{rcl}
\begin{aligned}
\sum_{i=0}^{N-1} &\frac{[2 \mu (\sigma_{u,i}^2+\delta) - \mu^2\sigma_{u,i}^2]}{(\sigma_{u,i}^2+\delta)^2} \text{E}\{e_{i,a}^2(\infty)\}  \\
&=\mu^2 \sum_{i=0}^{N-1} \frac{\sigma_{u,i}^2 ||\bm h_i||_2^2\sigma_v^2}{(\sigma_{u,i}^2+\delta)^2}  \\
& +2 \text{E}\left\lbrace \widetilde{\bm \psi}^\text{T}(\infty) \bm G^{-1}(\infty) P_{\mu \beta}(\bm \psi(\infty))\right\rbrace  \\
&+ \text{E} \left\lbrace P_{\mu \beta}(\bm \psi^\text{T}(\infty)) \bm G^{-1}(\infty) P_{\mu \beta}(\bm \psi(\infty))\right\rbrace.
\end{aligned}
\end{array}
\end{equation}

Under the steady-state,~\eqref{047} further becomes
\begin{equation}
\label{047x1}
\begin{array}{rcl}
\begin{aligned}
\sum_{i=0}^{N-1} &\frac{[2 \mu (\sigma_{u,i}^2+\delta) - \mu^2\sigma_{u,i}^2]}{(\sigma_{u,i}^2+\delta)^2} \text{E}\{e_{i,a}^2(\infty)\}  \\
 =&\mu^2 \sum_{i=0}^{N-1} \frac{\sigma_{u,i}^2 ||\bm h_i||_2^2\sigma_v^2}{(\sigma_{u,i}^2+\delta)^2} + 2 \text{E}\left\lbrace \widetilde{\bm w}^\text{T}(\infty) \bm G^{-1}(\infty) P_{\mu \beta}(\bm w(\infty))\right\rbrace \\
&+ \text{E} \left\lbrace P_{\mu \beta}(\bm w^\text{T}(\infty)) \bm G^{-1}(\infty) P_{\mu \beta}(\bm w(\infty))\right\rbrace.
\end{aligned}
\end{array}
\end{equation}

Taking advantage of $\text{E}\{e_{i,a}^2(\infty)\}\approx \text{MSD}(\infty)\sigma_{u,i}^2$ under the assumption of large enough $N$, we are able to derive $\text{MSD}(\infty)$ from~\eqref{047x1}:
\begin{equation}
\label{048}
\begin{array}{rcl}
\begin{aligned}
\text{MSD}(\infty) = \underbrace{\frac{1}{y} \mu^2 \sum_{i=0}^{N-1} \frac{\sigma_{u,i}^2 ||\bm h_i||_2^2\sigma_v^2}{(\sigma_{u,i}^2+\delta)^2}} \limits_{\text{PNSAF algorithm}} +\frac{1}{y} \Delta,
\end{aligned}
\end{array}
\end{equation}
where
\begin{equation}
\label{049}
\begin{array}{rcl}
\begin{aligned}
y = \sum_{i=0}^{N-1} \frac{[2 \mu (\sigma_{u,i}^2+\delta) - \mu^2\sigma_{u,i}^2]\sigma_{u,i}^2}{(\sigma_{u,i}^2+\delta)^2},
\end{aligned}
\end{array}
\end{equation}
\begin{equation}
\label{050}
\begin{array}{rcl}
\begin{aligned}
\Delta = & 2 \text{E}\left\lbrace \widetilde{\bm w}^\text{T}(\infty) \bm G^{-1}(\infty) P_{\mu \beta}(\bm w(\infty))\right\rbrace \\
&+ \text{E} \left\lbrace P_{\mu \beta}(\bm w(\infty)) \bm G^{-1}(\infty) P_{\mu \beta}(\bm w(\infty))\right\rbrace.
\end{aligned}
\end{array}
\end{equation}
Furthermore, $\text{EMSE}(\infty)$ can be calculated by~$\text{EMSE}(\infty) \approx \frac{1}{N} \sum_{i=0}^{N-1} \text{MSD}(\infty) \sigma_{u,i}^2$.

\textit{Remark~2:} When $\Delta=0$,~\eqref{048} will reduce to the $\text{MSD}(\infty)$ of the PNSAF algorithm, which shows that the proportionate matrix $G(k)$ does not affect the steady-state behavior of the PNSAF algorithm. That is to say, both NSAF and PNSAF algorithm have the same steady-state performance for the fixed step-size~$\mu$. Compared with the PNSAF algorithm, the $\text{MSD}(\infty)$ of the PFBS-PNSAF algorithm requires an additional term $\Delta$ resulting from the proximal step~\eqref{006x3}. Undoubtedly, the steady-state performance of the PFBS-PNSAF algorithm outperforms that of the PNSAF algorithm for sparse systems if, and only if $\Delta<0$ (whose possibility is clarified in Appendix~B). It is worth noting that $\beta$ should be chosen properly; otherwise, it will drive the proximal step~\eqref{006x3} improperly identifying Z and NZ coefficients in $\bm w^o$ at every iteration through the estimate $\bm \psi(k)$, thereby resulting in $\Delta>0$. In other words, there exists a range $0<\beta<\beta_\text{up}$ for $\Delta<0$ as be also seen in Fig.~\ref{Fig7}, where $\beta_\text{up}$ is unavailable despite the existence, as it requires knowing the true $\bm w^o$. As such, we will derive an adaptive scheme to choose~$\beta$ in the next section.
\begin{theorem}
    \textit{
    From $y>0$, we derive $0<\mu<2\frac{\sum_{i=0}^{N-1}\sigma_{u,i}^2+\delta}{\sum_{i=0}^{N-1}\sigma_{u,i}^2}$, which guarantees the mean square convergence of the PFBS-PNSAF algorithm.}
\end{theorem}

\textit{Remark~3:} From Theorems 1 and 2, the convergence condition of the PNSAF-type (including the PFBS-PNSAF) algorithms is expressed as
\begin{equation}
\label{051}
\begin{array}{rcl}
\begin{aligned}
0 < \mu< \min \left\lbrace 2\frac{\sum_{i=0}^{N-1}\sigma_{u,i}^2+\delta}{\sum_{i=0}^{N-1}\sigma_{u,i}^2}, \; \frac{2}{ \sum_{i=0}^{N-1} \frac{\lambda_\text{max}\left( \bar{\bm G}(k)\bm R_i \right)}{\sigma_{u,i}^2 + \delta} } \right\rbrace.
\end{aligned}
\end{array}
\end{equation}
For sufficiently large $N$, the above inequality is relaxed as
\begin{equation}
\label{052}
\begin{array}{rcl}
\begin{aligned}
0 < \mu< \min \left\lbrace 2\frac{\sum_{i=0}^{N-1}\sigma_{u,i}^2+\delta}{\sum_{i=0}^{N-1}\sigma_{u,i}^2}, \; \frac{2}{ \sum_{i=0}^{N-1} \frac{\sigma_{u,i}^2}{\sigma_{u,i}^2 + \delta} g_{\max}(k)} \right\rbrace,
\end{aligned}
\end{array}
\end{equation}
where $g_{\max}(k)\triangleq \lambda_\text{max}\left( \bar{\bm G}(k)\right)\leq 1$ due to $\text{Tr}\{ \bar{\bm G}(k) \}=1$. Most sparse systems are not extremely sparse that $\bm w^o$ has only a nonzero element so that $g_{\max}(k)$ is much less than~1, therefore, the convergence condition from~\eqref{052} can be simplified to
\begin{equation}
\label{053}
\begin{array}{rcl}
\begin{aligned}
0 &< \mu< 2\frac{\sum_{i=0}^{N-1}\sigma_{u,i}^2+\delta}{\sum_{i=0}^{N-1}\sigma_{u,i}^2} \approx 2 \text{ for small } \delta
\end{aligned}
\end{array}
\end{equation}
for the PNSAF-type algorithms in both mean and mean square senses. Moreover, as $\mu$ increases in this range, the steady-state behavior of the algorithm will deteriorate. On the other hand, $y$ can also characterize the convergence of the algorithm. Specifically, the fastest convergence is obtained when $y$ is minimum with respect to $\mu$, which leads to $\mu_{fast}=\frac{\sum_{i=0}^{N-1}\sigma_{u,i}^2+\delta}{\sum_{i=0}^{N-1}\sigma_{u,i}^2} \approx 1$. In other words, the increase of the step-size after larger than~$\mu_{fast}$ will slow down the algorithm convergence. As a consequence, we conclude the practical step-size range for the PNSAF-type algorithms:
\begin{equation}
\label{054}
\begin{array}{rcl}
\begin{aligned}
0 < \mu \leq \mu_{fast}.
\end{aligned}
\end{array}
\end{equation}
\section{Adaptation of $\beta$}
From Remark~2, it is known that the online choice of~$\beta$ is vital for the PFBS-PNSAF performance, but this is not easily derived from~\eqref{eq:21b} due to highly involved the nonlinear function on~$\beta$. To address this problem, as proved in~\cite{parikh2014proximal} the proximal operator can be approximated as
\begin{equation}
\label{059}
\begin{array}{rcl}
\begin{aligned}
\text{prox}_{\beta}^{F}(\bm \psi(k+1)) \approx \bm \psi(k+1) - \beta \triangledown F(\bm \psi(k+1)),
\end{aligned}
\end{array}
\end{equation}
when $\beta$ is small and $F(\cdot)$ is differentiable, where $\mu \beta$ is absorbed into $\beta$. Then, inspired by~our previous work in~\cite{yu2019sparsity}, the adaptation of~$\beta$ will be deployed. By subtracting~both sides of~\eqref{059} from~$\bm w^o$, we obtain the equation for the weights error vector:
\begin{equation}
\label{060}
\begin{array}{rcl}
\begin{aligned}
\widetilde{\bm w}(k+1) = \widetilde{\bm \psi}(k+1) + \beta \triangledown F(\bm \psi(k+1)).
\end{aligned}
\end{array}
\end{equation}

Taking the squared $l_2$-norm for both sides of \eqref{060} and enforcing the expectations over them, we obtain
\begin{equation}
\label{061}
\begin{array}{rcl}
\begin{aligned}
\text{E}&\{||\widetilde{\bm w}(k+1)||_2^2\} =  \text{E}\{||\widetilde{\bm \psi}(k+1)||_2^2\}  \\
&\;\;\;\;\;+ \beta \text{E}\{\widetilde{\bm \psi}^\text{T}(k+1) \triangledown F(\bm \psi(k+1))\} \\
&\;\;\;\;\;+ \beta^2 \text{E}\{||\triangledown F(\bm \psi(k+1))||_2^2\}.
\end{aligned}
\end{array}
\end{equation}

To reach the minimum of $\text{E}\{||\widetilde{\bm w}(k+1)||_2^2\}$, we set the derivative of~\eqref{061} with respect to $\beta$ to zero, yielding
\begin{equation}
\label{062}
\begin{array}{rcl}
\begin{aligned}
\beta_\text{opt}(k) = & -\frac{\text{E}\{\widetilde{\bm \psi}^\text{T}(k+1) \triangledown F(\bm \psi(k+1))\}}{\text{E}\{||\triangledown F(\bm \psi(k+1))||_2^2\}}. \\
\end{aligned}
\end{array}
\end{equation}

If $F(\cdot)$ is a real-valued convex function, from the definition of the sub-gradient~\cite{yu2019sparsity,chen2010regularized}
the following inequality will hold:
\begin{equation}
\label{063}
\begin{array}{rcl}
\begin{aligned}
\widetilde{\bm \psi}^\text{T}&(k+1) \triangledown F(\bm \psi(k+1)) \\
&= (\bm w^o - \bm \psi(k+1))^\text{T} \triangledown F(\bm \psi(k+1)) \\
&\leq  F(\bm w^o) -  F(\bm \psi(k+1)).
\end{aligned}
\end{array}
\end{equation}

Thus, using the above equation and approximating the expectations by their instantaneous values leads to
\begin{equation}
\label{064}
\begin{array}{rcl}
\begin{aligned}
\beta_\text{opt}(k) = \frac{\max[F(\bm \psi(k+1)) - F(\bm w^o),\; \tau]}{||\triangledown F(\bm \psi(k+1))||_2^2}. \\
\end{aligned}
\end{array}
\end{equation}
where $\tau \geq 0$ is a free parameter. It is evident that equation~\eqref{064} is not realistic as it requires knowing the true vector $\bm w^o$ beforehand to compute~$F(\bm w^o)$. To solve this problem, we provide a simple and effective way to estimate~$\bm w^o$, denoted as $\hat{\bm w}$:
\begin{equation}
\label{065}
\begin{array}{rcl}
\begin{aligned}
&\text{if}\;\text{mod}(k,M/N)=0\\
&\;\;\;\;\hat{\bm w} = \bm \psi(k+1)\\
&\text{else} \\
&\;\;\;\;\hat{\bm w} = 0.5\hat{\bm w} + 0.5\bm \psi(k+1)\\
&\text{end}.
\end{aligned}
\end{array}
\end{equation}
where $\text{mod}(a,b)$ takes the remainder of the division $a/b$. Subsequently, by combining~\eqref{064} and~\eqref{065}, the adaptation of $\beta$ is reformulated as
\begin{equation}
\label{066}
\begin{array}{rcl}
\begin{aligned}
\beta(k) = \frac{\max[F(\bm \psi(k+1)) - F(\hat{\bm w}),\; \tau]}{||\triangledown F(\bm \psi(k+1))||_2^2}. \\
\end{aligned}
\end{array}
\end{equation}

Finally, since $F(\bm \psi(k+1))=||\bm \psi(k+1)||_1$ is considered for the PFBS-PNSAF algorithm, it leads to $\triangledown F(\bm \psi(k+1)) = \text{sgn}(\bm \psi(k+1))$.

\textit{Remark~4:} Compared with the PNSAF counterpart, the PFBS-PNSAF algorithm requires additional calculations in terms of~(14), \eqref{065}, and \eqref{066}, thereby increasing the complexity of~$\mathcal{O}(M/N)$. However, the overall complexity of the PFBS-PNSAF algorithm with $\beta$-adaptation is still the order of~$\mathcal{O}(M)$ for each input sample.

\section{Simulations}
We present simulations to evaluate the proposed algorithm and the analysis results in the contexts of system identification and AEC. It is assumed that the length of the adaptive filter equals that of $\bm w^o$. The background noise $v(n)$ is zero-mean white Gaussian, giving rise to a signal-to-noise ratio (SNR) defined by $\text{SNR}=10\log10(\text{E}\{\bar{d}^2(n)\}/\sigma_v^2)$, where $\bar{d}(n)=\bm u^\text{T}(n)\bm w^o$. In the SAF structure, the analysis filters $\{H_i(z)\}_{i=0}^{N-1}$ are the cosine-modulated versions of a prototype lowpass filter. In our simulations, the prototype filter has lengths $L=17$, 33, and 65, respectively, for $N=2$, 4, 8 subbands, to guarantee 60 dB stopband attenuation.

\subsection{Verification of analyses for system identification}
In the sparse system identification, we consider two types of sparse systems: 1) TYPE-1, $\bm w^o$ has $M=128$ entries, where its $Q$ nonzero entries are Gaussian variables with zero mean and variance of $1/\sqrt{Q}$ and their positions are randomly selected from the binomial distribution; 2) TYPE-2, $\bm w^o$ is the network echo channel, i.e., model~1 from the ITU-T~G.168 standard~\cite{stnec2015}, of length $M=512$ with $Q=64$ nonzero entries. The input signal $u(n)$ is generated from a first-order autoregressive (AR) process $u(n)=0.8u(n-1)+\theta(n)$, where $\theta(n)$ is a white Gaussian noise with zero-mean and unit variance. For fairly evaluating the PNSAF and PFBS-PNSAF algorithms, we choose the proportionate rule given in~\eqref{002x3}, where choosing~$\epsilon=0.0001$ and $\zeta=0$. We use $10\log10(\text{MSD}(n))$ as a performance index, where the simulated MSD curves are the average of 100 independent trials. The steady-state MSDs are obtained by averaging 500 instantaneous MSD values in the steady-state. The regularization parameter in~\eqref{005} is chosen as~$\delta=0.001$.
\begin{figure}[htb]
    \centering
    \includegraphics[scale=0.45] {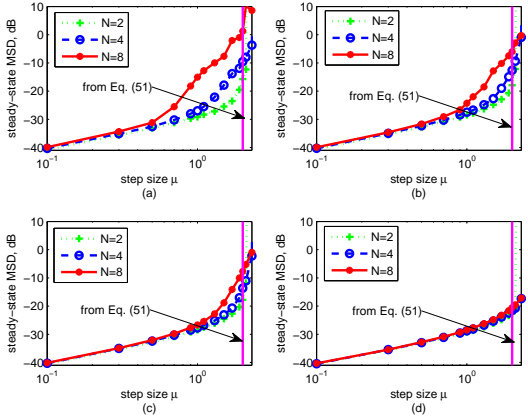}
    \vspace{-1em} \caption{Steady-state MSDs of the PNSAF algorithm. (a) TYPE-1: $Q=1$, (b) TYPE-1: $Q=4$, (c) TYPE-1: $Q=8$, (d)~TYPE-2. [SNR=30 dB] }
    \label{Fig2}
\end{figure}
\begin{figure}[htb]
    \centering
    \includegraphics[scale=0.45] {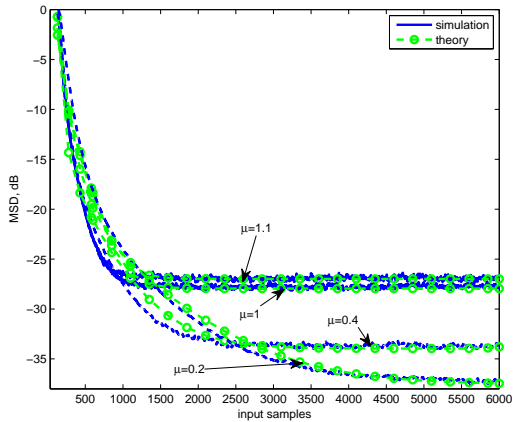}
    \vspace{-1em} \caption{MSD curves of the PNSAF algorithm. [$N=4$, SNR=30 dB, and TYPE-1: $Q=8$].}
    \label{Fig3}
\end{figure}

As stated in Remark~3, the convergence condition of the PFBS-PNSAF algorithm is the same as that of the PNSAF-type algorithms. Thus, Fig.~\ref{Fig2} shows the steady-state MSDs of the PNSAF algorithm as a function of~$\mu$. Only in an extremely sparse case such as Fig.~\ref{Fig2}(a), the stability range of the PNSAF-type algorithms is inversely proportional to $N$, since in this case, we know from~\eqref{052} the mean range will be narrower than the mean square range. However, realistic systems are not always extremely sparse so that the stability condition of the algorithm can be determined by~\eqref{053}, which does not depend on $N$ as shown in Fig.~\ref{Fig2}(d). Fig.~\ref{Fig3} shows the transient MSDs of the PNSAF algorithm for different step-sizes, where the theoretical curves are calculated by~\eqref{038} since this algorithm has no proximal step. As can be seen, as $\mu$ increases in the range~$(0,1]$, the convergence of the algorithm will become fast; however, when $\mu$ is larger than 1, the convergence rate will not be faster than that with $\mu=1$. This illustrates that the convergence condition~\eqref{054} is preferred for the PNSAF-type algorithms in practice.
\begin{figure}[htb]
    \centering
    \includegraphics[scale=0.45] {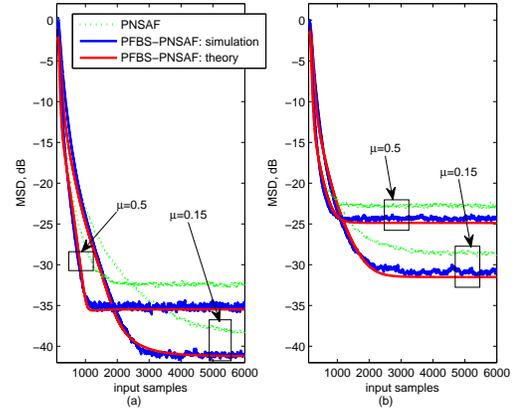}
    \vspace{-1em} \caption{MSD curves of the PNSAF and PFBS-PNSAF algorithms for TYPE-1: $Q=8$. (a) SNR=30 dB, (b) SNR=20 dB. [$N=4$]. In the PFBS-PNSAF algorithm, we set $\beta=9\times10^{-5}$ and $\beta=10\times10^{-5}$ for (a) and (b), respectively.}
    \label{Fig4}
\end{figure}

\begin{figure}[htb]
    \centering
    \includegraphics[scale=0.5] {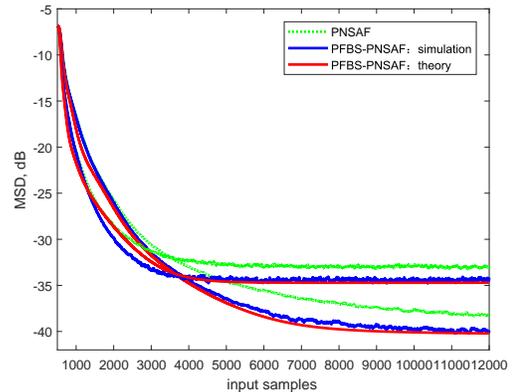}
    \vspace{-1em} \caption{MSD curves of the PNSAF and PFBS-PNSAF algorithms for TYPE-2. [$N=4$, SNR=30 dB]. We set $\beta=5\times10^{-6}$ for the PFBS-PNSAF algorithm.}
    \label{Fig6}
\end{figure}

In Fig.~\ref{Fig4}, we examine the transient MSD analysis for the PFBS-PNSAF algorithm, where the PNSAF algorithm is a comparison benchmark for identifying sparse system TYPE-1. Fig.~\ref{Fig6} depicts the transient MSD curves of the PFBS-PNSAF algorithm for identifying sparse system TYPE-2. As can be seen, in a sparse scenario, the PFBS-PNSAF algorithm is superior to the PNSAF algorithm in terms of the steady-state MSD performance, because the former has a proximal step to shrink most of the filter weights to zero. Moreover, for the PFBS-PNSAF algorithm, the fixed $\mu$ also controls the tradeoff between convergence rate and steady-state MSD. It can also be observed from~Figs.~\ref{Fig3}$\sim$\ref{Fig6} that theoretical results have almost good match with the simulated results. There is also the discrepancy between them, which mainly occurs in the case of larger step-size~$\mu=0.5$ or the transient of the PFBS-PNSAF algorithm, due to the adopted common assumptions to facilitate the analysis.

\begin{figure}[htb]
    \centering
    \includegraphics[scale=0.45] {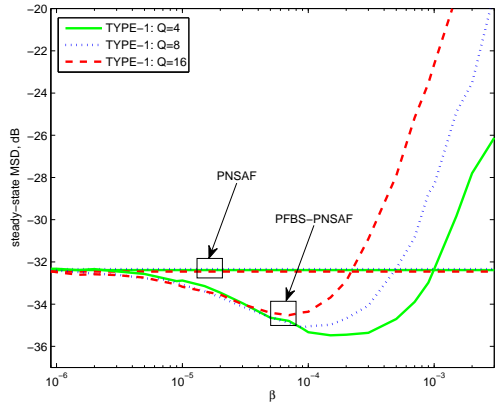}
    \vspace{-1em} \caption{Steady-state MSDs of the PFBS-PNSAF algorithm versus different $\beta$. [$N=4$, SNR=30 dB, and $\mu=0.5$].}
    \label{Fig7}
\end{figure}

Fig.~\ref{Fig7} investigates the effect of the thresholding parameter $\beta$ on the steady-state behavior of the PFBS-PNSAF algorithm. As analyzed in Remark~2, aiming to sparse systems, the PFBS-PNSAF algorithm will obtain better steady-state performance when $\beta$ is chosen in a certain range. As $Q$ is higher which tends to be less sparse, this range will become narrower.
\vspace{-1em}
\subsection{Comparison of algorithms in AEC}
\begin{figure}[htb]
    \centering
    \includegraphics[scale=0.40] {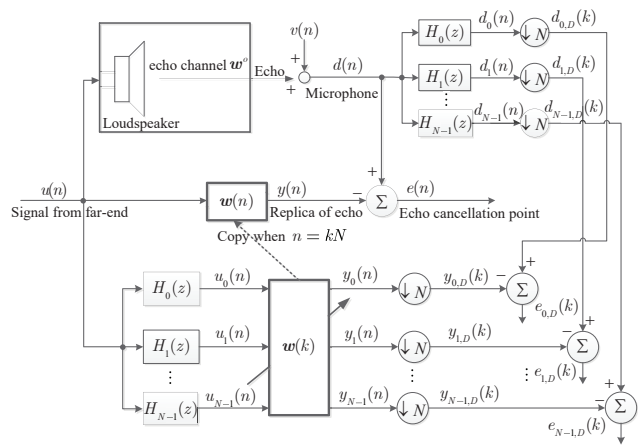}
    \vspace{-1em} \caption{Delayless diagram of multiband-structured SAF for AEC.}
    \label{Fig8}
\end{figure}

Fig.~\ref{Fig8} shows the Delayless diagram of multiband-structured SAF for AEC application, where $\bm w^o$ is the acoustic echo channel between loudspeaker and microphone. When the input signal $u(n)$ from the far-end is played at loudspeaker, through $\bm w^o$ the microphone will pick up the echo signal. The weights of the adaptive filter estimate $\bm w^o$, thus its output signal $y(n)= \bm u^\text{T}(n) \bm w(n)$ is the replica of the echo. Then, the echo can be canceled by subtracting $y(n)$ from $d(n)$, yielding the clean signal $e(n)=d(n)-y(n)$. It is worth noting that, here $e(n)$ is computed in an auxiliary loop, by copying $\bm w(k)$ to $\bm w(n)$ when $n=kN$. This avoids the signal delay problem in the original structure Fig.~\ref{Fig1} caused by the adopted analysis and synthesis filter banks. Importantly, the update equations of $\bm w(k)$ for SAF algorithms are the same in both structures. The acoustic echo channel $\bm w^o$ to be identified is from Fig.~3~in~\cite{zhao2014memory}. Also, $\bm w^o$ has a sudden change by shifting its~12 taps to the right at the middle of input samples, to assess the tracking performance of the algorithm.

\begin{figure}[htb]
    \centering
        \includegraphics[scale=0.5] {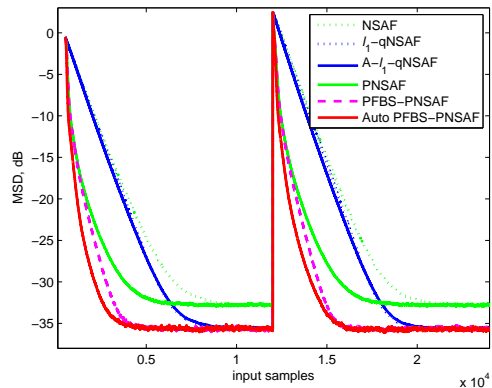}
    \vspace{-1em}
\caption{MSD curves of different subband algorithms for the AR input. [$N=4$, SNR=30 dB].
All the algorithms choose the same step-size $\mu=0.5$, and the proportionate parameters are chosen as $\epsilon=0.0001$ and $\zeta=-0.5$. Other parameters of algorithms are chosen as follows: $\mu=0.8$; $\beta=4\times10^{-6}$ ($l_1$-NSAF); $\delta_{\min}=0.001$ (A-$l_1$-NSAF); $\beta=5\times10^{-6}$ (PFBS-PNSAF), $\tau=0$ (Auto-PFBS-PNSAF). }
\label{Fig10a}
\end{figure}

Using the AR process in the previous subsection as the input signal, the proposed PFBS-PNSAF and Auto-PFBS-PNSAF (i.e., PFBS-PNSAF with adaptation of $\beta$) algorithms are compared with the NSAF, PNSAF, $l_1$-norm based quasi NSAF ($l_1$-qNSAF)~\cite{yu2016sparse}, and A-$l_1$-qNSAF (i.e., $l_1$-qNSAF with adaptation of $\beta$)~\cite{yu2016sparse} algorithms, and the MSD results are shown in Fig.~\ref{Fig10a}. Parameters of all the algorithms are tuned based on the same convergence or steady-state performance. The PFBS-PNSAF algorithm synthesizes the sparsity exploitation merits of both PNSAF and $l_1$-qNSAF algorithms, thus it has more outstanding performance in terms of convergence, steady-state, and tracking behaviors. Specifically, in the PFBS-PNSAF algorithm, the forward step drives the fast convergence due to the proportionate mechanism, and the proximal step further improves the steady-state performance due to attracting the majority of filter weights to zero. Similar to the $l_1$-qNSAF algorithm, the weak point of the PFBS-PNSAF algorithm is also that the best $\beta$ is often chosen in a trial and error way. Fortunately, by adaptively adjusting $\beta$ derived from the minimum MSD principle, the Auto-PFBS-PNSAF algorithm avoids the parameter problem.

\begin{figure}[htb]
    \centering
    \includegraphics[scale=0.5] {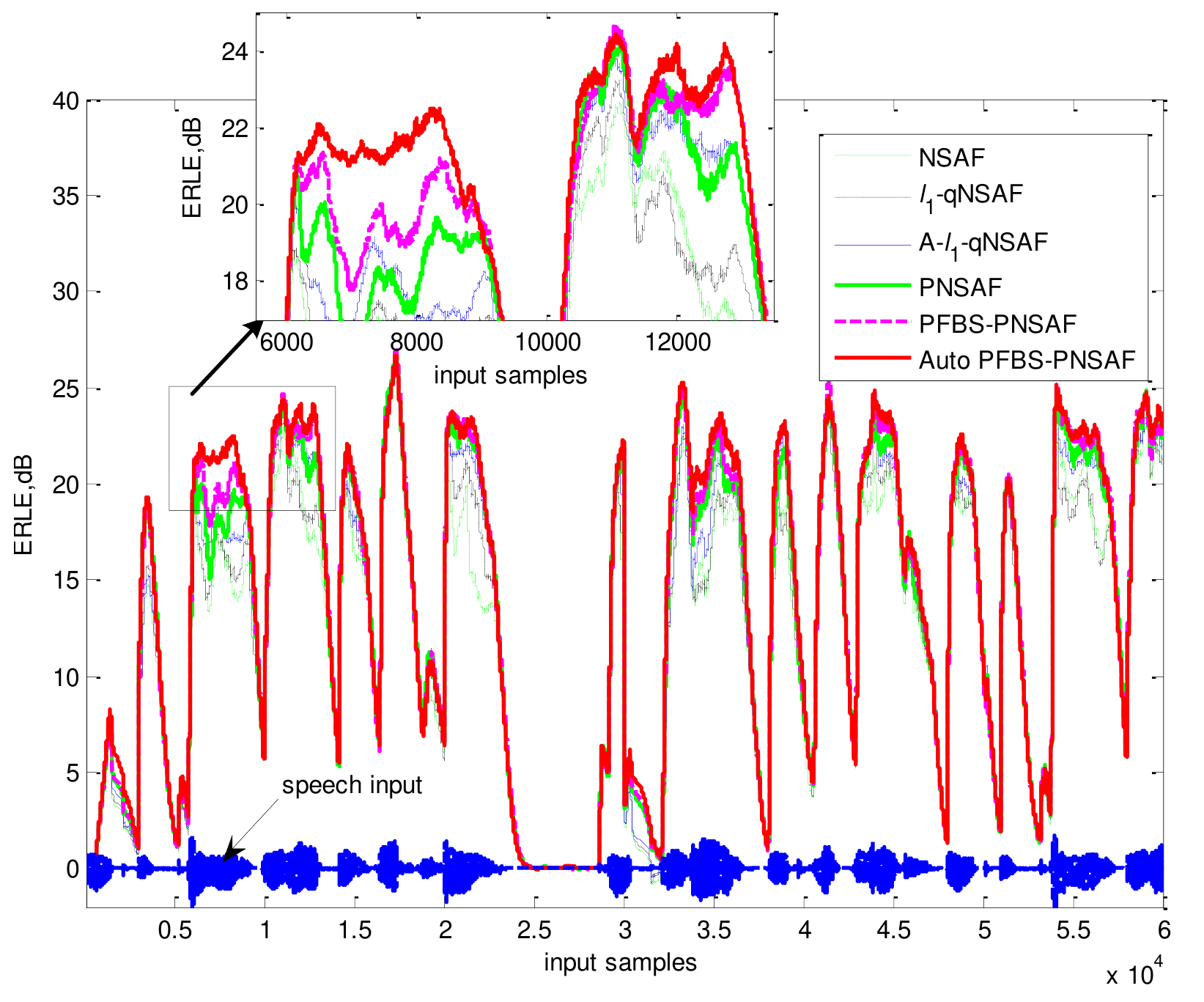}
    \vspace{-1em} \caption{ERLE curves of different subband algorithms for the speech input. [$N=8$, SNR=30 dB, and single run].}
    \label{Fig12}
\end{figure}

Furthermore, we compare these algorithms in Fig.~\ref{Fig12} by using a realistic speech as the input signal. In this scenario, we use the echo return loss enhancement (ERLE) as a performance index~\cite{yu2019m}, defined as $\text{ERLE}(n) = 10\log_{10} (\text{avg} \{d^2(n)\}/\text{avg} \{e^2(n)\})$, where $\text{avg}(\cdot)$ is a smooth filtering in the form $\sigma_d^2(n)=\chi\sigma_d^2(n) + (1-\chi)d^2(n)$ with $\chi=0.996$. When using the speech input, to prevent the division by zero like in~\eqref{005}, we set the regularization parameter $\delta$, i.e., $\delta=20\sigma_u^2/N$ (NSAF-type) and $\delta=20\sigma_u^2/M$ (PNSAF-type), where $\sigma_u^2$ is the power of speech signal. Parameters setting of algorithms is the same as in~Fig.~\ref{Fig10a}. It is clear that the proposed PFBS-PNSAF and Auto-PFBS-PNSAF algorithms exhibit faster convergence and higher ERLE than the other algorithms, which means they provide a better talk quality. The Auto-PFBS-PNSAF algorithm is preferred as it does not require the choice of $\beta$.

\begin{figure}[htb]
    \centering
    \includegraphics[scale=0.5] {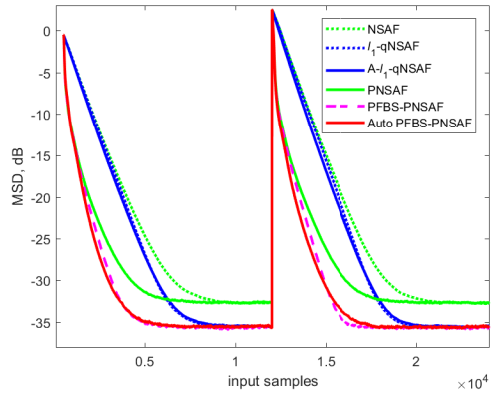}
    \vspace{-1em} \caption{MSD curves of different algorithms for the AR input. [$N=4$ and SNR=30 dB].}
    \label{Fig13}
\end{figure}

\begin{figure}[htb]
    \centering
    \includegraphics[scale=0.52] {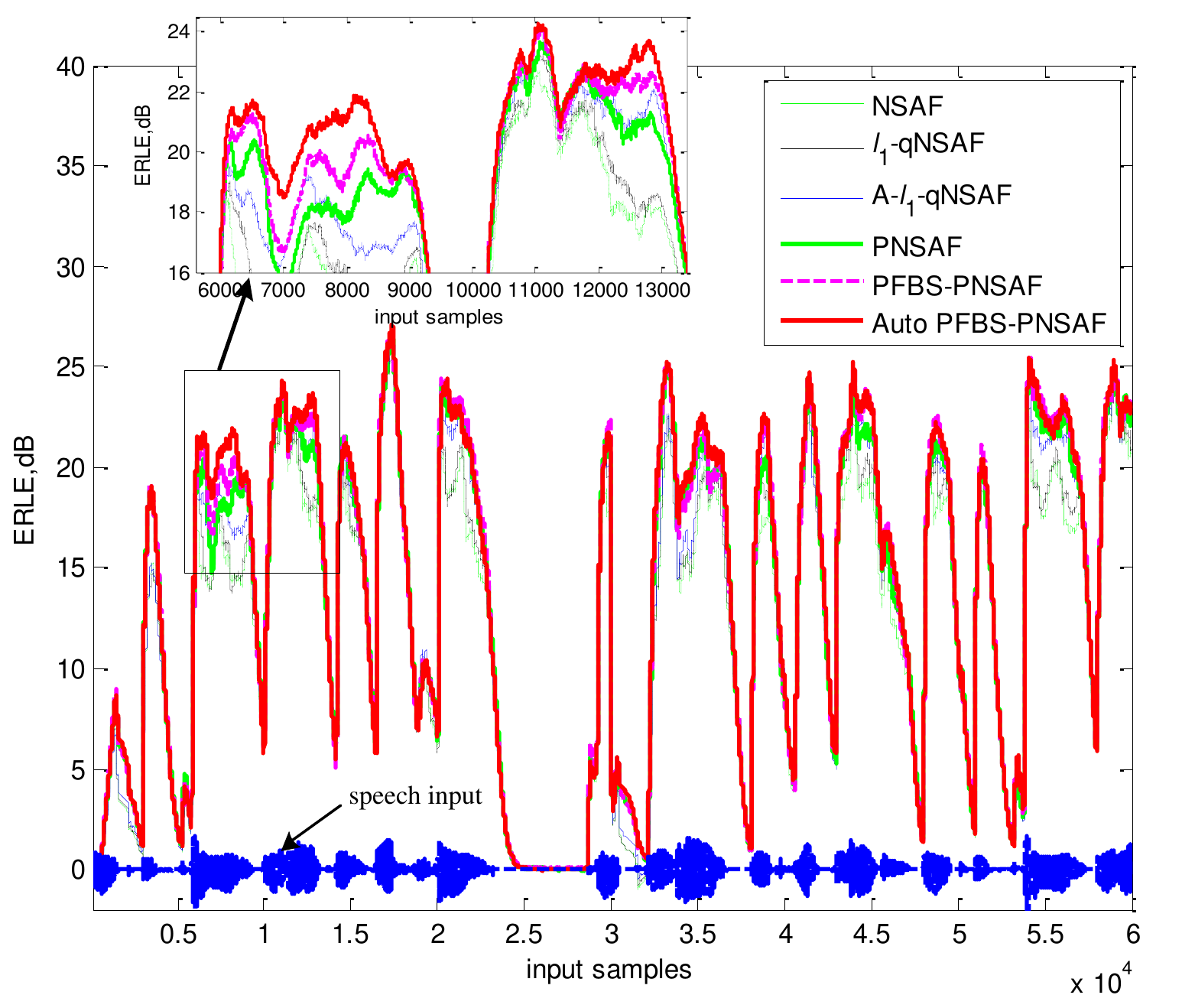}
    \vspace{-1em} \caption{ERLE curves of different subband algorithms for the speech input. [$N=8$, SNR=30 dB, and single run].}
    \label{Fig14}
\end{figure}

Without loss of generality, by applying another proportionate strategy instead of~\eqref{002x3}, i.e., the one that developed firstly in the proportionate NLMS algorithm~\cite{duttweiler2000proportionate}, we again evaluate the proposed PFBS-PNSAF algorithm in Figs.~\ref{Fig13} and~\ref{Fig14}. For all the algorithms of PNSAF-type, the proportionate factors are computed as,
\begin{equation}
\label{0r61}
\begin{array}{rcl}
\begin{aligned}
q_m(k) &= \max \left[  \rho \max \left( \gamma, \{|w_m(k)|\}_{m=1}^{M}\right) , |w_m(k)| \right],\\
g_m(k) &= \frac{q_m(k)}{\sum_{m=1}^M q_m(k)},
\end{aligned}
\end{array}
\end{equation}
where the parameter $\rho$ (typical range 0.01$\sim$0.05) avoids the freeze of the filter's weights $\{w_m(k)\}_{m=1}^M$ when their absolute values are much smaller than the largest one, and $\gamma$ with a typical value of 0.01 is to allow the adaptation even if $\bm w(0) = \bm 0$ at initialization. The PNSAF algorithm with~\eqref{0r61} was also introduced in~\cite{abadi2011family}. Figs.~\ref{Fig13} and~\ref{Fig14} depict the MSD and ERLE results of the algorithms for the AR input and the speech input, respectively. In two figures, we only set the parameters in~\eqref{0r61} to $\rho=0.04$ and $\gamma=0.01$, and other parameters of algorithms are the same as before. As expected, in spite of using the proportionate rule given in~\eqref{0r61}, the proposed PFBS-PNSAF algorithm outperforms its NSAF counterparts in either convergence or steady-state performance. Because of the adaptive adjustment of the thresholding parameter~$\beta$, the proposed 'Auto' variant has also more robust performance.


\begin{figure}[htb]
    \centering
    \includegraphics[scale=0.5] {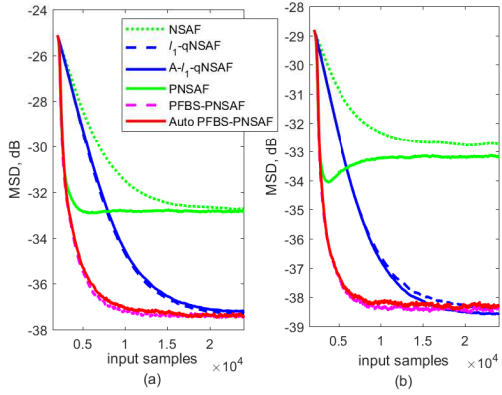}
    \vspace{-1em} \caption{MSD curves of different algorithms. (a) $T_s=160$ ms, (b) $T_s=400$ ms. [AR input, $N=4$ and SNR=30 dB]. Some parameters of algorithms are re-tuned as follows: $\beta=1\times10^{-6}$ for $l_1$-NSAF; $\delta_{\min}=0.002$ for A-$l_1$-NSAF; $\beta=1\times10^{-6}$ for PFBS-PNSAF, $\tau=0.001$ for Auto-PFBS-PNSAF. Other parameters are as those in Fig.~\ref{Fig13} before.}
    \label{Fig18}
\end{figure}

Finally, the proposed algorithm is evaluated, for identifying acoustic impulse responses with different reverberation time. Impulse responses are generated by~\cite{Alien1976Image}, with length of~$M=2048$. Correspondingly, the results of the algorithms are shown in~Fig.~\ref{Fig18}. As one can see, the proposed PFBS-PNSAF algorithm still achieves good performance in contrast with the previous counterparts.

\section{Conclusions}
Based on PFBS and soft-thresholding techniques, we have integrated the benefits of the proportionate and sparsity-aware ideas characterizing the underlying sparsity of the systems, to derive the PFBS-PNSAF algorithm, and further provided its delayless implementation applied to AEC. By employing some commonly used assumptions, the mean and mean-square performances of this algorithm are studied in detail. Moreover, to equip the algorithm with robustness against the choice of a thresholding parameter, we also proposed an adaptive method for choosing it. Simulation results in various environments have demonstrated the effectiveness of our algorithms and the theoretical analysis.

\vspace{-1em}
\appendices
\numberwithin{equation}{section}
\section{Calculation of the term $(b)$ in~\eqref{036}}
\renewcommand{\appendixname}{Appendix~}
For the term~(b) in~\eqref{036}, we can divide it into two parts when $i=j$ and $i\neq j$, namely,
\begin{equation}
\label{0A1}
\begin{array}{rcl}
\begin{aligned}
&\text{E}\left\lbrace \sum_{i=0}^{N-1} \frac{\bm u_i(k) \bm u_i^\text{T}(k) }{\sigma_{u,i}^2 + \delta} \widetilde{\bm W}(k) \sum_{j=0}^{N-1} \frac{\bm u_j(k) \bm u_j^\text{T}(k)}{\sigma_{u,j}^2 + \delta}\right\rbrace \\
& = \text{E}\left\lbrace \sum_{i=0}^{N-1} \frac{\bm u_i(k) \bm u_i^\text{T}(k) }{\sigma_{u,i}^2 + \delta} \widetilde{\bm W}(k) \frac{\bm u_i(k) \bm u_i^\text{T}(k)}{\sigma_{u,i}^2 + \delta}\right\rbrace \\
&+\text{E}\left\lbrace \sum_{i=0}^{N-1} \frac{\bm u_i(k) \bm u_i^\text{T}(k) }{\sigma_{u,i}^2 + \delta} \widetilde{\bm W}(k) \sum\limits_{j=0,j\neq i}^{N-1} \frac{\bm u_j(k) \bm u_j^\text{T}(k)}{\sigma_{u,j}^2 + \delta}\right\rbrace.
\end{aligned}
\end{array}
\end{equation}

By performing the Gaussian moment factoring theorem~\cite{yin2010stochastic}, the first term at the right side of~\eqref{0A1} is deduced to
\begin{equation}
\label{0A2}
\begin{array}{rcl}
\begin{aligned}
&\text{E}\left\lbrace \frac{\bm u_i(k) \bm u_i^\text{T}(k) }{\sigma_{u,i}^2 + \delta} \widetilde{\bm W}(k) \frac{\bm u_i(k) \bm u_i^\text{T}(k)}{\sigma_{u,i}^2 + \delta}\right\rbrace  \\
&= 2 \text{E}\left\lbrace \frac{\bm u_i(k) \bm u_i^\text{T}(k) }{\sigma_{u,i}^2 + \delta}\right\rbrace \widetilde{\bm W}(k) \text{E}\left\lbrace \frac{\bm u_i(k) \bm u_i^\text{T}(k)}{\sigma_{u,i}^2 + \delta}\right\rbrace \\
&\;\;\;\;+ \text{E}\left\lbrace \frac{\bm u_i(k) \bm u_i^\text{T}(k) }{\sigma_{u,i}^2 + \delta}\right\rbrace \text{Tr}\left( \widetilde{\bm W}(k) \text{E}\left\lbrace \frac{\bm u_i(k) \bm u_i^\text{T}(k) }{\sigma_{u,i}^2 + \delta}\right\rbrace \right) \\
&= \frac{1}{(\sigma_{u,i}^2 + \delta)^2} \left[ 2\bm R_i \widetilde{\bm W}(k) \bm R_i + \bm R_i \text{Tr}\left( \widetilde{\bm W}(k) \bm R_i \right) \right].
\end{aligned}
\end{array}
\end{equation}
where $\text{Tr}(\cdot)$ represents the trace of a matrix. For the last term of~\eqref{0A1}, since the
different subband vectors $\bm u_i(k)$ and $\bm u_j(k)$ are weakly correlated~\cite{lee2009subband}, we can assume their correlation as zero. Then, by combining~\eqref{0A1} and \eqref{0A2}, we arrive at:
\begin{equation}
\label{0A3}
\begin{array}{rcl}
\begin{aligned}
&\text{E}\left\lbrace \sum_{i=0}^{N-1} \frac{\bm u_i(k) \bm u_i^\text{T}(k) }{\sigma_{u,i}^2 + \delta} \widetilde{\bm W}(k) \sum_{j=0}^{N-1} \frac{\bm u_j(k) \bm u_j^\text{T}(k)}{\sigma_{u,j}^2 + \delta}\right\rbrace
\\&= \sum_{i=0}^{N-1} \frac{1}{(\sigma_{u,i}^2 + \delta)^2} \left[\bm R_i \widetilde{\bm W}(k) \bm R_i + \bm R_i \text{Tr}\left( \widetilde{\bm W}(k) \bm R_i \right) \right].
\end{aligned}
\end{array}
\end{equation}

\section{Proof of $\Delta<0$}
Based on assumption 4, we can rewritten~$\Delta$ in~\eqref{050} as
\begin{equation}
\label{0B1}
\begin{array}{rcl}
\begin{aligned}
\Delta = &2 \sum_{m=0}^{M} \bar{g}_m^{-1}(\infty) \text{E}\{ \widetilde{w}_m(\infty) P_{\mu \beta}(w_m(\infty))\} \\
&+ \sum_{m=0}^{M} \bar{g}_m^{-1}(\infty) \text{E}\{ P_{\mu \beta}^2(w_m(\infty))\},
\end{aligned}
\end{array}
\end{equation}
where $\delta$ is negligible for brevity due to small value. Again recalling~$\bm R_i \approx \sigma_{u,i}^2$ under the assumption of large enough $N$,~\eqref{033} can be rewritten in the component-wise form as
\begin{equation}
\label{0B2}
\begin{array}{rcl}
\begin{aligned}
\mathrm{E}\{\widetilde{w}_m(\infty)\} = \left\{ \begin{aligned}
&\beta N^{-1} \bar{g}_m^{-1}(\infty)  \times \\
&\;\;\mathrm{E}\{\mathrm{sgn}(w_m(\infty))\}, \mathrm{ if }\;|w_m(\infty)| > \mu\beta,\\
&w_m^o, \mathrm{ if }\; |w_m(\infty)| \leq \mu\beta,
\end{aligned} \right.
\end{aligned}
\end{array}
\end{equation}
where $m=1,...,M$.

To proceed, we will classify the components of the sparse vector $\bm w^o$ into two categories: the sets of zero and nonzero entries, denoted by Z and NZ, respectively, i.e., $w_m^o=0$ for $m \in \text{Z}$ and
$w_m^o \neq 0$ for $m \in \text{NZ}$. It is assumed that $\mu$ and $\beta$ are appropriately small, $P_{\mu \beta}(w_m(\infty))$ given in~\eqref{026} can distinguish correctly Z (when $|w_m|<\mu \beta$) and NZ (when $|w_m|>\mu \beta$) entires. By doing so,~\eqref{0B2} can be changed to
\begin{equation}
\label{0B3}
\text{E}\{\widetilde{w}_m(\infty)\} = \left\{ \begin{aligned}
&\beta N^{-1} \bar{g}_m^{-1}(\infty) \text{sgn}(w_m^o), \text{ if } m\in \text{NZ}, \\
&0, \text{ if } m\in \text{Z}.
\end{aligned} \right.
\end{equation}

Similarly, we also have
\begin{equation}
\label{0B4}
\begin{array}{rcl}
\begin{aligned}
&\text{E}\{ \widetilde{w}_m(\infty) P_{\mu \beta}(w_m(\infty))\} \\
&\;\;\;\;\;\;\;= \left\{ \begin{aligned}
&\mu \beta \text{E}\{\widetilde{w}_m(\infty)\} \text{sgn}(w_m^o) , \text{ if } m\in \text{NZ}, \\
&-\text{E}\{\widetilde{w}_m^2(\infty)\} , \text{ if } m\in \text{Z},
\end{aligned} \right.
\end{aligned}
\end{array}
\end{equation}

\begin{equation}
\label{0B5}
\text{E}\{P_{\mu \beta}^2(w_m(\infty))\} = \left\{ \begin{aligned}
&\mu^2 \beta^2, \text{ if } m\in \text{NZ}, \\
&\text{E}\{\widetilde{w}_m^2(\infty)\} , \text{ if } m\in \text{Z}.
\end{aligned} \right.
\end{equation}

By substituting~\eqref{0B3}$\sim$\eqref{0B5} into \eqref{0B1}, we obtain the following relation
\begin{equation}
\label{0B6}
\begin{array}{rcl}
\begin{aligned}
\Delta =& \sum_{m \in \text{NZ}} \mu \beta^2 \bar{g}_m^{-2}(\infty) (2N^{-1} + \mu )  \\
&- \sum_{m \in \text{Z}} \bar{g}_m^{-1}(\infty) \text{E}\{\widetilde{w}_m^2(\infty)\}.
\end{aligned}
\end{array}
\end{equation}

Thanks to $\bm w^o$ being sparse and $\sum_{m=1}^M \bar{g}_m(\infty)=1$, $\bar{g}_m^{-1}(\infty)$ for $m \in \text{Z}$ is larger than $\bar{g}_m^{-1}(\infty)$ for $m \in \text{NZ}$. For instance, in an extreme sparse case that $\bm w^o$ has only a nonzero element, thus from~\eqref{002x3} with $\zeta=0$ we have $\bar{g}_m^{-1}(\infty)=2M$ for $m \in \text{Z}$ and $\bar{g}_m^{-1}(\infty)=\frac{2M}{M+1}\approx 2$ for $m \in \text{NZ}$. In addition, $\beta$ is very small as shown in simulations. As such, the second term on the right side of~\eqref{0B6} would be larger than the first term. Consequently, $\Delta$ is likely to be true when $\bm w^o$ is sparse. Also note that $\Delta<0$ is not possible for non-sparse~$\bm w^o$.

\ifCLASSOPTIONcaptionsoff
  \newpage
\fi

\bibliographystyle{IEEEtran}
\bibliography{IEEEabrv,mybibfile}

\begin{thebibliography}{10}
\providecommand{\url}[1]{#1}
\csname url@samestyle\endcsname
\providecommand{\newblock}{\relax}
\providecommand{\bibinfo}[2]{#2}
\providecommand{\BIBentrySTDinterwordspacing}{\spaceskip=0pt\relax}
\providecommand{\BIBentryALTinterwordstretchfactor}{4}
\providecommand{\BIBentryALTinterwordspacing}{\spaceskip=\fontdimen2\font plus
\BIBentryALTinterwordstretchfactor\fontdimen3\font minus
  \fontdimen4\font\relax}
\providecommand{\BIBforeignlanguage}[2]{{%
\expandafter\ifx\csname l@#1\endcsname\relax
\typeout{** WARNING: IEEEtran.bst: No hyphenation pattern has been}%
\typeout{** loaded for the language `#1'. Using the pattern for}%
\typeout{** the default language instead.}%
\else
\language=\csname l@#1\endcsname
\fi
#2}}
\providecommand{\BIBdecl}{\relax}
\BIBdecl

\bibitem{chen2016generalized}
B.~Chen, L.~Xing, H.~Zhao, N.~Zheng, and J.~C. Principe, ``Generalized
  correntropy for robust adaptive filtering,'' \emph{IEEE Transactions on
  Signal Processing}, vol.~64, no.~13, pp. 3376--3387, 2016.

\bibitem{chen2017kernel}
B.~Chen, L.~Xing, B.~Xu, H.~Zhao, N.~Zheng, and J.~C. Principe, ``Kernel
  risk-sensitive loss: definition, properties and application to robust
  adaptive filtering,'' \emph{IEEE Transactions on Signal Processing}, vol.~65,
  no.~11, pp. 2888--2901, 2017.

\bibitem{lee2009subband}
K.-A. Lee, W.-S. Gan, and S.~M. Kuo, \emph{Subband adaptive filtering: theory
  and implementation}.\hskip 1em plus 0.5em minus 0.4em\relax John Wiley \&
  Sons, 2009.

\bibitem{pradhan2017improved}
S.~Pradhan, V.~Patel, D.~Somani, and N.~V. George, ``An improved proportionate
  delayless multiband-structured subband adaptive feedback canceller for
  digital hearing aids,'' \emph{IEEE/ACM Transactions on Audio, Speech, and
  Language Processing}, vol.~25, no.~8, pp. 1633--1643, 2017.

\bibitem{intadap}
R.~de~Lamare and R.~Sampaio-Neto, ``Adaptive reduced-rank mmse filtering with
  interpolated fir filters and adaptive interpolators,'' \emph{IEEE Signal
  Processing Letters}, vol.~12, no.~3, pp. 177--180, 2005.

\bibitem{jio}
R.~C. de~Lamare and R.~Sampaio-Neto, ``Reduced-rank adaptive filtering based on
  joint iterative optimization of adaptive filters,'' \emph{IEEE Signal
  Processing Letters}, vol.~14, no.~12, pp. 980--983, 2007.

\bibitem{jidf}
R.~C. de~Lamare and R.~Sampaio-Neto, ``Adaptive reduced-rank processing based
  on joint and iterative interpolation, decimation, and filtering,'' \emph{IEEE
  Transactions on Signal Processing}, vol.~57, no.~7, pp. 2503--2514, 2009.

\bibitem{sjidf}
R.~Fa, R.~C. de~Lamare, and L.~Wang, ``Reduced-rank stap schemes for airborne
  radar based on switched joint interpolation, decimation and filtering
  algorithm,'' \emph{IEEE Transactions on Signal Processing}, vol.~58, no.~8,
  pp. 4182--4194, 2010.

\bibitem{jiols}
R.~C. de~Lamare and R.~Sampaio-Neto, ``Reduced-rank space time adaptive
  interference suppression with joint iterative least squares algorithms for
  spread-spectrum systems,'' \emph{IEEE Transactions on Vehicular Technology},
  vol.~59, no.~3, pp. 1217--1228, 2010.

\bibitem{jiomimo}
R.~C. de~Lamare and R.~Sampaio-Neto, ``Adaptive reduced-rank equalization
  algorithms based on alternating optimization design techniques for mimo
  systems,'' \emph{IEEE Transactions on Vehicular Technology}, vol.~60, no.~6,
  pp. 2482--2494, 2011.

\bibitem{wlmwf}
N.~Song, R.~C. de~Lamare, M.~Haardt, and M.~Wolf, ``Adaptive widely linear
  reduced-rank interference suppression based on the multistage wiener
  filter,'' \emph{IEEE Transactions on Signal Processing}, vol.~60, no.~8, pp.
  4003--4016, 2012.

\bibitem{wljio}
N.~Song, W.~U. Alokozai, R.~C. de~Lamare, and M.~Haardt, ``Adaptive widely
  linear reduced-rank beamforming based on joint iterative optimization,''
  \emph{IEEE Signal Processing Letters}, vol.~21, no.~3, pp. 265--269, 2014.

\bibitem{jiodoa}
L.~Wang, R.~C. de~Lamare, and M.~Haardt, ``Direction finding algorithms based
  on joint iterative subspace optimization,'' \emph{IEEE Transactions on
  Aerospace and Electronic Systems}, vol.~50, no.~4, pp. 2541--2553, 2014.

\bibitem{barc}
R.~C. de~Lamare, R.~Sampaio-Neto, and M.~Haardt, ``Blind adaptive constrained
  constant-modulus reduced-rank interference suppression algorithms based on
  interpolation and switched decimation,'' \emph{IEEE Transactions on Signal
  Processing}, vol.~59, no.~2, pp. 681--695, 2011.

\bibitem{rrbf}
S.~Li, R.~C. de~Lamare, and R.~Fa, ``Reduced-rank linear interference
  suppression for ds-uwb systems based on switched approximations of adaptive
  basis functions,'' \emph{IEEE Transactions on Vehicular Technology}, vol.~60,
  no.~2, pp. 485--497, 2011.

\bibitem{rrser}
Y.~Cai, R.~C. de~Lamare, B.~Champagne, B.~Qin, and M.~Zhao, ``Adaptive
  reduced-rank receive processing based on minimum symbol-error-rate criterion
  for large-scale multiple-antenna systems,'' \emph{IEEE Transactions on
  Communications}, vol.~63, no.~11, pp. 4185--4201, 2015.

\bibitem{rralr}
L.~Qiu, Y.~Cai, R.~C. de~Lamare, and M.~Zhao, ``Reduced-rank doa estimation
  algorithms based on alternating low-rank decomposition,'' \emph{IEEE Signal
  Processing Letters}, vol.~23, no.~5, pp. 565--569, 2016.

\bibitem{l1stap}
Z.~Yang, R.~C. de~Lamare, and X.~Li, ``<formula formulatype="inline"><tex
  notation="tex">$l_1$</tex> </formula>-regularized stap algorithms with a
  generalized sidelobe canceler architecture for airborne radar,'' \emph{IEEE
  Transactions on Signal Processing}, vol.~60, no.~2, pp. 674--686, 2012.

\bibitem{smtvb}
R.~C. de~Lamare and P.~S.~R. Diniz, ``Set-membership adaptive algorithms based
  on time-varying error bounds for cdma interference suppression,'' \emph{IEEE
  Transactions on Vehicular Technology}, vol.~58, no.~2, pp. 644--654, 2009.

\bibitem{smce}
T.~Wang, R.~C. de~Lamare, and P.~D. Mitchell, ``Low-complexity set-membership
  channel estimation for cooperative wireless sensor networks,'' \emph{IEEE
  Transactions on Vehicular Technology}, vol.~60, no.~6, pp. 2594--2607, 2011.

\bibitem{dce}
S.~Xu, R.~C. de~Lamare, and H.~V. Poor, ``Distributed compressed estimation
  based on compressive sensing,'' \emph{IEEE Signal Processing Letters},
  vol.~22, no.~9, pp. 1311--1315, 2015.

\bibitem{damdc}
T.~G. Miller, S.~Xu, R.~C. de~Lamare, and H.~V. Poor, ``Distributed spectrum
  estimation based on alternating mixed discrete-continuous adaptation,''
  \emph{IEEE Signal Processing Letters}, vol.~23, no.~4, pp. 551--555, 2016.

\bibitem{arc}
F.~G. Almeida~Neto, R.~C. De~Lamare, V.~H. Nascimento, and Y.~V. Zakharov,
  ``Adaptive reweighting homotopy algorithms applied to beamforming,''
  \emph{IEEE Transactions on Aerospace and Electronic Systems}, vol.~51, no.~3,
  pp. 1902--1915, 2015.

\bibitem{spa}
R.~C. De~Lamare and R.~Sampaio-Neto, ``Minimum mean-squared error iterative
  successive parallel arbitrated decision feedback detectors for ds-cdma
  systems,'' \emph{IEEE Transactions on Communications}, vol.~56, no.~5, pp.
  778--789, 2008.

\bibitem{mbdf}
R.~C. de~Lamare, ``Adaptive and iterative multi-branch mmse decision feedback
  detection algorithms for multi-antenna systems,'' \emph{IEEE Transactions on
  Wireless Communications}, vol.~12, no.~10, pp. 5294--5308, 2013.

\bibitem{bfidd}
A.~G.~D. Uchoa, C.~T. Healy, and R.~C. de~Lamare, ``Iterative detection and
  decoding algorithms for mimo systems in block-fading channels using ldpc
  codes,'' \emph{IEEE Transactions on Vehicular Technology}, vol.~65, no.~4,
  pp. 2735--2741, 2016.

\bibitem{aaidd}
R.~B. Di~Renna and R.~C. de~Lamare, ``Adaptive activity-aware iterative
  detection for massive machine-type communications,'' \emph{IEEE Wireless
  Communications Letters}, vol.~8, no.~6, pp. 1631--1634, 2019.

\bibitem{listmtc}
R.~B. Di~Renna and R.~C. de~Lamare, ``Iterative list detection and decoding for
  massive machine-type communications,'' \emph{IEEE Transactions on
  Communications}, vol.~68, no.~10, pp. 6276--6288, 2020.

\bibitem{msgampmtc}
R.~B.~D. Renna and R.~C. de~Lamare, ``Dynamic message scheduling based on
  activity-aware residual belief propagation for asynchronous mmtc,''
  \emph{IEEE Wireless Communications Letters}, vol.~10, no.~6, pp. 1290--1294,
  2021.

\bibitem{dynovs}
Z.~Shao, L.~T.~N. Landau, and R.~C. de~Lamare, ``Dynamic oversampling for 1-bit
  adcs in large-scale multiple-antenna systems,'' \emph{IEEE Transactions on
  Communications}, vol.~69, no.~5, pp. 3423--3435, 2021.

\bibitem{locsme}
H.~Ruan and R.~C. de~Lamare, ``Robust adaptive beamforming using a
  low-complexity shrinkage-based mismatch estimation algorithm,'' \emph{IEEE
  Signal Processing Letters}, vol.~21, no.~1, pp. 60--64, 2014.

\bibitem{okspme}
H.~Ruan and R.~C. de~Lamare, ``Robust adaptive beamforming based on low-rank
  and cross-correlation techniques,'' \emph{IEEE Transactions on Signal
  Processing}, vol.~64, no.~15, pp. 3919--3932, 2016.

\bibitem{lrcc}
H.~Ruan and R.~C. de~Lamare, ``Distributed robust beamforming based on low-rank
  and cross-correlation techniques: Design and analysis,'' \emph{IEEE
  Transactions on Signal Processing}, vol.~67, no.~24, pp. 6411--6423, 2019.

\bibitem{rcoprime}
X.~Wang, Z.~Yang, J.~Huang, and R.~C. de~Lamare, ``Robust two-stage
  reduced-dimension sparsity-aware stap for airborne radar with coprime
  arrays,'' \emph{IEEE Transactions on Signal Processing}, vol.~68, pp. 81--96,
  2020.

\bibitem{sayed2003fundamentals}
A.~H. Sayed, \emph{Fundamentals of adaptive filtering}.\hskip 1em plus 0.5em
  minus 0.4em\relax John Wiley \& Sons, 2003.

\bibitem{yang2018comparative}
F.~Yang and J.~Yang, ``A comparative survey of fast affine projection
  algorithms,'' \emph{Digital Signal Processing}, vol.~83, pp. 297--322, 2018.

\bibitem{zakharov2008low}
Y.~V. Zakharov, G.~P. White, and J.~Liu, ``Low-complexity {RLS} algorithms
  using dichotomous coordinate descent iterations,'' \emph{IEEE Transactions on
  Signal Processing}, vol.~56, no.~7, pp. 3150--3161, 2008.

\bibitem{lee2004improving}
K.-A. Lee and W.-S. Gan, ``Improving convergence of the {NLMS} algorithm using
  constrained subband updates,'' \emph{IEEE signal processing letters},
  vol.~11, no.~9, pp. 736--739, 2004.

\bibitem{lee2007delayless}
K.-A. Lee and W.-S. Gan, ``On delayless architecture for the normalized subband
  adaptive filter,'' in \emph{2007 IEEE International Conference on Multimedia
  and Expo}, 2007, pp. 1595--1598.

\bibitem{ni2009variable}
J.~Ni and F.~Li, ``A variable step-size matrix normalized subband adaptive
  filter,'' \emph{IEEE Transactions on Audio, Speech, and Language Processing},
  vol.~18, no.~6, pp. 1290--1299, 2009.

\bibitem{seo2014variable}
J.-H. Seo and P.~Park, ``Variable individual step-size subband adaptive
  filtering algorithm,'' \emph{Electronics letters}, vol.~50, no.~3, pp.
  177--178, 2014.

\bibitem{ni2010adaptive}
J.~Ni and F.~Li, ``Adaptive combination of subband adaptive filters for
  acoustic echo cancellation,'' \emph{IEEE Transactions on Consumer
  Electronics}, vol.~56, no.~3, pp. 1549--1555, 2010.

\bibitem{yang2012improved}
F.~Yang, M.~Wu, P.~Ji, and J.~Yang, ``An improved multiband-structured subband
  adaptive filter algorithm,'' \emph{IEEE Signal Processing Letters}, vol.~19,
  no.~10, pp. 647--650, 2012.

\bibitem{yang2015low}
F.~Yang, M.~Wu, P.~Ji, and J.~Yang, ``Low-complexity implementation of the
  improved multiband-structured subband adaptive filter algorithm,'' \emph{IEEE
  Transactions on Signal Processing}, vol.~63, no.~19, pp. 5133--5148, 2015.

\bibitem{radecki2002echo}
J.~Radecki, Z.~Zilic, and K.~Radecka, ``Echo cancellation in {IP} networks,''
  in \emph{The 2002 45th Midwest Symposium on Circuits and Systems, 2002.
  MWSCAS-2002.}, vol.~2, 2002, pp. II--II.

\bibitem{yukawa2008efficient}
M.~Yukawa, R.~C. De~Lamare, and R.~Sampaio-Neto, ``Efficient acoustic echo
  cancellation with reduced-rank adaptive filtering based on selective
  decimation and adaptive interpolation,'' \emph{IEEE Transactions on Audio,
  Speech, and Language Processing}, vol.~16, no.~4, pp. 696--710, 2008.

\bibitem{schreiber1995advanced}
W.~F. Schreiber, ``Advanced television systems for terrestrial broadcasting:
  Some problems and some proposed solutions,'' \emph{Proceedings of the IEEE},
  vol.~83, no.~6, pp. 958--981, 1995.

\bibitem{loganathan2009class}
P.~Loganathan, A.~W. Khong, and P.~A. Naylor, ``A class of
  sparseness-controlled algorithms for echo cancellation,'' \emph{IEEE
  Transactions on Audio, Speech, and Language Processing}, vol.~17, no.~8, pp.
  1591--1601, 2009.

\bibitem{benesty2002improved}
J.~Benesty and S.~L. Gay, ``An improved {PNLMS} algorithm,'' in \emph{2002 IEEE
  International Conference on Acoustics, Speech, and Signal Processing},
  vol.~2, 2002, pp. II--1881.

\bibitem{duttweiler2000proportionate}
D.~L. Duttweiler, ``Proportionate normalized least-mean-squares adaptation in
  echo cancelers,'' \emph{IEEE Transactions on speech and audio processing},
  vol.~8, no.~5, pp. 508--518, 2000.

\bibitem{abadi2009proportionate}
M.~S.~E. Abadi, ``Proportionate normalized subband adaptive filter algorithms
  for sparse system identification,'' \emph{Signal Processing}, vol.~89, no.~7,
  pp. 1467--1474, 2009.

\bibitem{abadi2011family}
M.~S.~E. Abadi and S.~Kadkhodazadeh, ``A family of proportionate normalized
  subband adaptive filter algorithms,'' \emph{Journal of the Franklin
  Institute}, vol. 348, no.~2, pp. 212--238, 2011.

\bibitem{baraniuk2007compressive}
R.~G. Baraniuk, ``Compressive sensing [lecture notes],'' \emph{IEEE signal
  processing magazine}, vol.~24, no.~4, pp. 118--121, 2007.

\bibitem{gu2009l}
Y.~Gu, J.~Jin, and S.~Mei, ``$l_0$ norm constraint {LMS} algorithm for sparse
  system identification,'' \emph{IEEE Signal Processing Letters}, vol.~16,
  no.~9, pp. 774--777, 2009.

\bibitem{de2014sparsity}
R.~C. de~Lamare and R.~Sampaio-Neto, ``Sparsity-aware adaptive algorithms based
  on alternating optimization and shrinkage,'' \emph{IEEE Signal Processing
  Letters}, vol.~21, no.~2, pp. 225--229, 2014.

\bibitem{yu2016sparse}
Y.~Yu, H.~Zhao, and B.~Chen, ``Sparse normalized subband adaptive filter
  algorithm with $l_0$-norm constraint,'' \emph{Journal of the Franklin
  Institute}, vol. 353, no.~18, pp. 5121--5136, 2016.

\bibitem{yu2019sparsity}
Y.~Yu, H.~Zhao, R.~C. de~Lamare, and L.~Lu, ``Sparsity-aware subband adaptive
  algorithms with adjustable penalties,'' \emph{Digital Signal Processing},
  vol.~84, pp. 93--106, 2019.

\bibitem{pelekanakis2012new}
K.~Pelekanakis and M.~Chitre, ``New sparse adaptive algorithms based on the
  natural gradient and the $l_0$-norm,'' \emph{IEEE Journal of Oceanic
  Engineering}, vol.~38, no.~2, pp. 323--332, 2012.

\bibitem{das2016improving}
R.~L. Das and M.~Chakraborty, ``Improving the performance of the {PNLMS}
  algorithm using $l_1$-norm regularization,'' \emph{IEEE/ACM Transactions on
  Audio, Speech, and Language Processing}, vol.~24, no.~7, pp. 1280--1290,
  2016.

\bibitem{jin2017enhanced}
Z.~Jin, Y.~Li, and Y.~Wang, ``An enhanced set-membership {PNLMS} algorithm with
  a correntropy induced metric constraint for acoustic channel estimation,''
  \emph{Entropy}, vol.~19, no.~6, p. 281, 2017.

\bibitem{ferreira2016low}
T.~N. Ferreira, M.~V. Lima, P.~S. Diniz, and W.~A. Martins, ``Low-complexity
  proportionate algorithms with sparsity-promoting penalties,'' in \emph{2016
  IEEE International Symposium on Circuits and Systems (ISCAS)}.\hskip 1em plus
  0.5em minus 0.4em\relax IEEE, 2016, pp. 253--256.

\bibitem{yamagishi2011acceleration}
M.~Yamagishi, M.~Yukawa, and I.~Yamada, ``Acceleration of adaptive proximal
  forward-backward splitting method and its application to sparse system
  identification,'' in \emph{2011 IEEE International Conference on Acoustics,
  Speech and Signal Processing (ICASSP)}.\hskip 1em plus 0.5em minus
  0.4em\relax IEEE, 2011, pp. 4296--4299.

\bibitem{jeong2018automatic}
K.~Jeong, M.~Yukawa, M.~Yamagishi, and I.~Yamada, ``Automatic shrinkage tuning
  robust to input correlation for sparsity-aware adaptive filtering,'' in
  \emph{2018 IEEE International Conference on Acoustics, Speech and Signal
  Processing (ICASSP)}.\hskip 1em plus 0.5em minus 0.4em\relax IEEE, 2018, pp.
  4314--4318.

\bibitem{zheng2017robust}
Z.~Zheng, Z.~Liu, H.~Zhao, Y.~Yu, and L.~Lu, ``Robust set-membership normalized
  subband adaptive filtering algorithms and their application to acoustic echo
  cancellation,'' \emph{IEEE Transactions on Circuits and Systems I: Regular
  Papers}, vol.~64, no.~8, pp. 2098--2111, 2017.

\bibitem{parikh2014proximal}
N.~Parikh, S.~Boyd \emph{et~al.}, ``Proximal algorithms,'' \emph{Foundations
  and Trends{\textregistered} in Optimization}, vol.~1, no.~3, pp. 127--239,
  2014.

\bibitem{hu2014sparse}
T.~Hu and D.~B. Chklovskii, ``Sparse {LMS} via online linearized bregman
  iteration,'' in \emph{IEEE International Conference on Acoustics, Speech and
  Signal Processing (ICASSP)}.\hskip 1em plus 0.5em minus 0.4em\relax IEEE,
  2014, pp. 7213--7217.

\bibitem{lunglmayr2016efficient}
M.~Lunglmayr and M.~Huemer, ``Efficient linearized bregman iteration for sparse
  adaptive filters and kaczmarz solvers,'' in \emph{IEEE Sensor Array and
  Multichannel Signal Processing Workshop (SAM)}.\hskip 1em plus 0.5em minus
  0.4em\relax IEEE, 2016, pp. 1--5.

\bibitem{lunglmayr2017scaled}
M.~Lunglmayr, B.~Hiptmair, and M.~Huemer, ``Scaled linearized bregman
  iterations for fixed point implementation,'' in \emph{IEEE International
  Symposium on Circuits and Systems (ISCAS)}.\hskip 1em plus 0.5em minus
  0.4em\relax IEEE, 2017, pp. 1--4.

\bibitem{chen2014steady}
B.~Chen, L.~Xing, J.~Liang, N.~Zheng, and J.~C. Principe, ``Steady-state
  mean-square error analysis for adaptive filtering under the maximum
  correntropy criterion,'' \emph{IEEE Signal Processing Letters}, vol.~21,
  no.~7, pp. 880--884, 2014.

\bibitem{dang2019kernel}
L.~Dang, B.~Chen, S.~Wang, Y.~Gu, and J.~C. Pr{\'\i}ncipe, ``Kernel kalman
  filtering with conditional embedding and maximum correntropy criterion,''
  \emph{IEEE Transactions on Circuits and Systems I: Regular Papers}, vol.~66,
  no.~11, pp. 4265--4277, 2019.

\bibitem{yin2010stochastic}
W.~Yin and A.~S. Mehr, ``Stochastic analysis of the normalized subband adaptive
  filter algorithm,'' \emph{IEEE Transactions on Circuits and Systems I:
  Regular Papers}, vol.~58, no.~5, pp. 1020--1033, 2010.

\bibitem{jeong2015mean}
J.~J. Jeong, S.~H. Kim, G.~Koo, and S.~W. Kim, ``Mean-square deviation analysis
  of multiband-structured subband adaptive filter algorithm,'' \emph{IEEE
  Transactions on Signal Processing}, vol.~64, no.~4, pp. 985--994, 2015.

\bibitem{loganathan2010performance}
P.~Loganathan, E.~A. Habets, and P.~A. Naylor, ``Performance analysis of
  {IPNLMS} for identification of time-varying systems,'' in \emph{2010 IEEE
  International Conference on Acoustics, Speech and Signal Processing}.\hskip
  1em plus 0.5em minus 0.4em\relax IEEE, 2010, pp. 317--320.

\bibitem{haddad2014transient}
D.~B. Haddad and M.~R. Petraglia, ``Transient and steady-state {MSE} analysis
  of the {IMPNLMS} algorithm,'' \emph{Digital Signal Processing}, vol.~33, pp.
  50--59, 2014.

\bibitem{zhang2018mean}
S.~Zhang and W.~X. Zheng, ``Mean-square analysis of multi-sampled
  multiband-structured subband filtering algorithm,'' \emph{IEEE Transactions
  on Circuits and Systems I: Regular Papers}, vol.~66, no.~3, pp. 1051--1062,
  2018.

\bibitem{chen2010regularized}
Y.~Chen, Y.~Gu, and A.~O. Hero, ``Regularized least-mean-square algorithms,''
  \emph{arXiv preprint arXiv:1012.5066}, 2010.

\bibitem{stnec2015}
\emph{Digital Network Echo Cancellers Recommendation}, Std. ITU-TG.168 (V8),
  2015.

\bibitem{zhao2014memory}
H.~Zhao, Y.~Yu, S.~Gao, X.~Zeng, and Z.~He, ``Memory proportionate {APA} with
  individual activation factors for acoustic echo cancellation,''
  \emph{IEEE/ACM transactions on audio, speech, and language processing},
  vol.~22, no.~6, pp. 1047--1055, 2014.

\bibitem{yu2019m}
Y.~Yu, H.~He, B.~Chen, J.~Li, Y.~Zhang, and L.~Lu, ``{M}-estimate based
  normalized subband adaptive filter algorithm: Performance analysis and
  improvements,'' \emph{IEEE/ACM Transactions on Audio, Speech, and Language
  Processing}, vol.~28, pp. 225--239, 2020.

\bibitem{Alien1976Image}
Alien, J., and B., ``Image method for efficiently simulating small-room
  acoustics,'' \emph{The Journal of the Acoustical Society of America},
  vol.~60, no.~S1, p.~S9, 1976.

\end{thebibliography}

\end{document}